\documentstyle[prd,floats,aps]{revtex}
\input psfig
\pssilent
\def\ut#1{#1\llap{\lower2ex\hbox{$\widetilde{\hphantom{#1}}$}}}
\begin{document}
\input psfig \pssilent 
\title{Canonical quantum gravity in the Vassiliev invariants arena:\\
I. Kinematical structure}
\author{
Cayetano Di Bartolo$^{1}$
Rodolfo Gambini$^{2}$, Jorge Griego$^2$, Jorge Pullin$^3$}
\address{1. Departamento de F\'{\i}sica, Universidad Sim\'on Bol\'{\i}var,\\
Aptdo. 89000, Caracas 1080-A, Venezuela.}
\address{2. Instituto de F\'{\i}sica, Facultad de Ciencias, 
Igu\'a 4225, esq. Mataojo, Montevideo, Uruguay.}
\address{ 3. Center for Gravitational Physics and Geometry, 
Department of Physics,\\ The
Pennsylvania State University, 104 Davey Lab, University Park, PA
16802.} 

\maketitle
\begin{abstract}
We generalize the idea of Vassiliev invariants to the spin network
context, with the aim of using these invariants as a kinematical arena
for a canonical quantization of gravity.  This paper presents a
detailed construction of these invariants (both ambient and regular
isotopic) requiring a significant elaboration based on the use of
Chern-Simons perturbation theory which extends the work of Kauffman,
Martin and Witten to four-valent networks. We show that this space of
knot invariants has the crucial property ---from the point of view of
the quantization of gravity--- of being loop differentiable in the
sense of distributions. This allows the definition of  diffeomorphism
and Hamiltonian constraints. We show that the invariants are
annihilated by the diffeomorphism constraint. In a companion paper we
elaborate on the definition of a Hamiltonian constraint, discuss the
constraint algebra, and show that the construction leads to a
consistent theory of canonical quantum gravity.
\end{abstract}
\vspace{-10cm} 
\begin{flushright}
\baselineskip=15pt
CGPG-99/11-2  \\
gr-qc/9911009\\
\end{flushright}
\vspace{7.5cm}

\section{Introduction}
\subsection{Preliminaries}
Since the early attempts in the 60's to construct a canonical quantum
theory of general relativity, a persistent problem has been the
definition of well-defined quantum operatorial expressions for the
constraints of the theory. In particular, the realization of the
quantum Hamiltonian constraint (the Wheeler--DeWitt equation), given
its nonlinear structure in terms of momenta, has been particularly
elusive. In terms of the usual canonical variables (metric and
extrinsic curvature), the first impediment found was that the
constraint is highly non-polynomial. The introduction of the Ashtekar
variables \cite{As86} seemed to provide a natural way to construct a
polynomial expression for the Hamiltonian constraint. However, the
formalism presented two significant complications: a) the variables in
question were complex; b) the polynomial Hamiltonian constraint was a
density of weight two. The latter fact is crucial at the time of
promoting the constraint to a quantum operator. Since in a manifold
without a prescribed classical metric (as is the case in quantum
gravity where the metric is a quantum operator) the only naturally
defined density is of weight one (the Dirac delta), most attempts to
regularize \cite{RoSm90,Ga91,BrPu93,RoSm94} the Hamiltonian constraint
produced operators that exhibited dependence on external, artificial
metric structures introduced via the regularization procedure. Such
remnant dependences on fiducial metric structures almost inevitably
imply that the constraint algebra will not be properly implemented at
a quantum level. Since fiducial structures do not transform
appropriately under the action of the diffeomorphism constraint, the
net effect is to make the Hamiltonian constraint not covariant. At
most one could hope for formal consistency, when the algebra was
computed ignoring regulators \cite{GaGaPu}.

An important step forward towards solving the two problems we
mentioned above was a construction due to Thiemann \cite{QSD1}.  He
realized that it was indeed possible to construct a polynomial version
of the single-densitized Hamiltonian constraint in terms of
Ashtekar-like variables. To begin with, Thiemann uses the real
connection variables introduced by Barbero \cite{Ba}, thus bypassing
the issues associated with the reality conditions of the traditional
Ashtekar formulation. Barbero's variables consist of an $SU(2)$
connection $A_a^i$ and a set of triads $\tilde{E}^a_i$ (throughout
this paper we denote density weights with tildes).  The basic idea to
polynomialize the Hamiltonian constraint is the key canonical
identity,
\begin{equation}
{2 \over G} \{A_a^i,V\} = {\ut{\epsilon}
\raise2ex\hbox{$$}\,{}_{abc} \tilde{E}^b_j \tilde{E}^c_k
\epsilon^{ijk} \over  \det{q}},
\end{equation}
where $G$ is Newton's constant, and, \begin{equation}
V = \int d^3x 
\sqrt{\ut{\epsilon}
\raise2ex\hbox{$$}\,{}_{abc} \tilde{E}^a_i \tilde{E}^b_j \tilde{E}^c_k
\epsilon^{ijk}},
\end{equation}
is the volume element. In terms of these expressions the
singly-densitized Hamiltonian constraint can be written as a
polynomial operator. To simplify the discussion we focus on
the Hamiltonian of Euclidean general relativity (Thiemann
\cite{QSD1} has shown that an analogous construction is possible
for the Hamiltonian of the Lorentzian theory, which involves an
extra term),
\begin{equation}
\tilde{H} = {\tilde{E}^a_j \tilde{E}^a_k 
\epsilon^{ijk} \over  \det{q}} F_{ab}^i, \label{classh}
\end{equation}
which can therefore be written as,
\begin{equation}
\tilde{H} = {2 \over G} \{A_a^i,V\} F_{bc}^i \tilde{\epsilon}^{abc},
\end{equation}
In \cite{QSD1}, Thiemann elaborated a proposal for promoting this
expression to a quantum operator acting on diffeomorphism-invariant
wavefunctions associated with cylindrical functions of the
connection (see \cite{AsMaMo} for a quick, physical, description).
The latter are functions constructed by considering gauge invariant
quantities parameterized by multivalently-intersecting graphs
embedded in three dimensions, usually referred to in this context
as spin networks.  On the space of cylindrical functions there are
known measures \cite{AsLe} that make states parameterized by
non-diffeomorphically-related spin networks orthogonal. The
proposed Hamiltonian constraint is strictly speaking only defined
on the diffeomorphism-invariant dual of the space of cylindrical
functions constructed using the measure we just referred to.

Thiemann's proposal fulfilled several of the usual expected
requirements for a quantized Hamiltonian constraint of general
relativity. The operator was well defined and finite (even if one
coupled the theory to matter), could be made self-adjoint in a
controlled fashion under the above mentioned measure, and was
covariant under finite diffeomorphisms in a detailed sense. Because
the operator was realized in a space of diffeomorphism-invariant
wavefunctions, the expected result for the commutator of two
Hamiltonian constraints is zero, and indeed this is the result that
is obtained. In this approach there is no implementation of the
infinitesimal generator of diffeomorphisms. Technically speaking,
Thiemann's construction constitutes one of the first-ever
consistent theories of quantum gravity. Consistency is a
prerequisite for any physically relevant theory, but is not a
sufficient condition. It is still to be understood if Thiemann's
proposal embodies the correct physics of general relativity at a
semiclassical level.

A shadow of doubt was cast on the physical relevance of Thiemann's
construction through an argument constructed by Lewandowski and Marolf
\cite{LeMa} (see also \cite{GaLeMaPu} for further
elaboration). Lewandowski and Marolf considered a different space of
wavefunctions on which to define a Hamiltonian constraint. In
particular, the new ``habitat'' (to use their terminology) contains
the diffeomorphism-invariant dual of cylindrical functions we
considered above, but also contains non-diffeomorphism invariant
functions. Better yet, the same construction one followed to promote
the Hamiltonian constraint to an operator in Thiemann's proposal is
applicable in the new habitat. However, a crucial difference that
stems from the correspondence with the classical Poisson algebra of
constraints, is that the Hamiltonian one constructs should not commute
with itself, but yield a commutator that is proportional to a
diffeomorphism (in the new habitat the diffeomorphism constraint is a
well defined operator yielding the natural geometric action). Yet, the
Hamiltonian one obtains applying Thiemann's proposal commutes with
itself, even in the habitat of non-diffeomorphism invariant states. In
fact, it was also shown that several modifications of Thiemann's
construction also produce Hamiltonians with vanishing
commutators. In addition, if one artificially modified the Hamiltonians to
yield a non-vanishing commutator, the end result did not appear to correspond
to the quantization of the right hand side of the classical commutator 
\cite{GaLeMaPu}.

Notice that these arguments do not {\em prove} that Thiemann's
construction is incorrect in the diffeomorphism invariant context. It
could still be the case that in that context the theory yields the
correct semi-classical limit. In fact, Thiemann's construction in
$2+1$ dimensions seems to have the same difficulties as in $3+1$
dimensions, and still contains within its space of solutions those of
the more traditional quantizations of $2+1$ gravity\cite{QSD4}. By
carefully choosing an inner product and imposing other restrictions,
the approach can be made to yield only the usual solutions of the
$2+1$ theory. However, it is disturbing that one does not seem to find
a context where this construction reproduces the classical Poisson
algebra with a non-vanishing commutator of two Hamiltonians. The
commutativity might also have implications for the semiclassical limit
of the theory \cite{Sm96}.

In a separate set of developments, that has its roots in formal
calculations performed some time ago in the language of loops and
involving the doubly-densitized Hamiltonian
\cite{BrGaPu93,BrGaPunpb,GaPubook,FoGaPu}, another space of
wavefunctions has been considered. These are the wavefunctions that
arise when one considers in the space of gauge invariant functions of
a connection the measure given by the exponential of the Chern--Simons
action. The attractiveness of these wavefunctions stems from a large
body of results in Chern--Simons theory that relates these states to
well studied knot invariants (see \cite{Gu93} for references),
including the Vassiliev invariants.  In addition, the Chern--Simons
measure is in a formal sense a solution of the Hamiltonian constraint
of quantum gravity with a cosmological constant
\cite{Ko,BrGaPunpb}. Moreover, the space is quite distinct from that
of cylindrical functions in the sense that the Vassiliev invariants
take non-trivial values on infinite sets of loops (or spin networks)
whereas cylindrical functions take non-trivial values on finite sets
of spin networks. Most of the literature on these invariants, however,
has been focusing on the case of loops, whereas the context that is
of interest for quantum gravity requires the use of spin networks.
Part of the problems we had to confront were to suitably generalize
the Vassiliev invariants to the spin network context. Partial progress
along these lines, but with a different motivation, was achieved by
Kauffman and Lins \cite{KaLi} and Witten and Martin \cite{Ma}, but we
still need to elaborate on this to have an appropriate setting for
doing quantum gravity.

Why might it be interesting to consider these states in 
addressing the difficulties we
discussed that appear in relation to Thiemann's Hamiltonian?
Essentially because in both the space considered by Thiemann and in
the habitat of Lewandowski and Marolf, gauge invariant operators
involving $\hat{F}_{ab}$ do not have a natural action. In spaces of
these types, the usual way to represent such an operator would be to
use the loop derivative, as is customarily in the non-diffeomorphism
invariant Yang--Mills context. However, it was soon recognized that a
conflict arises between the use of the loop derivative and the
diffeomorphism invariance of the spaces in question. To put it
naively, the action of the loop derivative ``appends a loop of
infinitesimal area'', and in a diffeomorphism invariant context there
is no meaning to ``infinitesimal area'' in terms of functions of loops
or spin nets. In the limit involved in the definition of the loop
derivative, the numerator is either zero or finite and the denominator
goes to zero. This is what one expects when one is considering what is
tantamount to a ``derivative of a step function''. The result should
be a distribution. Finding the correct distribution that represents
the operator is the main problem. In the new space we are proposing,
it turns out that the loop derivative has a well defined
distributional action, which can be derived from the definition of the
invariants in terms of the Chern-Simons formulae. 

Another advantage of the kind of states we are considering is that
generically they have support on an infinite number of spin
networks. The loop representation dual of cylindrical functions has
support on a finite number of spin nets only. This might have physical
consequences. For instance, if one considers the computation of the
volume of a finite region, the result will generically vanish for the
loop representation non-diffeomorphism-invariant duals of the
cylindrical functions (unless the volume in question contains at least
one four or higher valent vertex of the finite set of spin networks
characterizing the state). For the Vassiliev invariants, the result of
the action of the volume operator of a finite region is generically
non-vanishing. Thiemann already notices this drawback of cylindrical
functions in the $2+1$ context \cite{QSD4}. Essentially, using
Vassiliev invariants is tantamount to considering infinite
superpositions of cylindrical states. To give an intuitive (and
imprecise) analogy, working with cylindrical functions is like working
with Dirac deltas in one-dimensional calculus. Working with the
Vassiliev invariants is more alike to working with ordinary smooth
functions. We will see in the companion paper that these requirements
seem to lead to fruitful developments at least in the context of $2+1$
dimensional gravity.

Having at hand a definition for the loop derivative in this
context, one is in a much better position to achieve a more
realistic quantum operator representing the Hamiltonian constraint.
To illustrate this point, let us recall briefly Thiemann's
construction. It starts from the classical Hamiltonian
(\ref{classh}) which, introducing a simplicial decomposition in
space $\Delta$, can be written as,
\begin{equation}
H = \epsilon^{ijk} {\rm Tr}( h_{\alpha_{ij}} h^{-1}_{e_k(\Delta)}
\{ h_{e_k(\Delta)},V\}).
\end{equation}
where $h_{\alpha_{ij}}$ is a holonomy along one of the triangles of the
tetrahedra of the simplicial decomposition and $e_k$ are edges of the
triangles. This expression obviously represents the Hamiltonian
(\ref{classh}) in the limit in which the tetrahedra of the
discretization become infinitesimal, the holonomy along the closed
loop becoming $F_{ab}$ times the tangent vectors of the edges of the
triangle. This quantity is later promoted to a quantum operator,
essentially by adding hats in all the quantities involved. The
resulting operator acts in the space of diffeomorphism invariant
cylindrical functions. An important point is that in this context
there is no notion of shrinking the simplicial decomposition to
infinitesimal size. In fact, this feature allows to bypass in the
construction the evaluation of an ill defined limit (the one appearing
in the loop derivative) and to simply consider a {\em finite} holonomy
as representing $F_{ab}$. These steps are technically correct, but one
can roughly expect that if one is representing $F_{ab}$ by a finite
holonomy instead of an infinitesimal one, one would encounter
difficulties at the time of computing commutators. A finite holonomy
differs from an infinitesimal one by an infinite number of higher
order terms, that in general will have non-vanishing contributions to
commutators. 

\subsection{Strategy}

In this paper we will discuss two main topics: a) the generalization
of the notion of Vassiliev invariant to the spin network context; this will
give a concrete footing to the wavefunctions we will consider in the
companion paper for the action of the Hamiltonian constraint; b)
the definition of a loop derivative and an infinitesimal generator of
diffeomorhpisms in this arena, which are key to defining the constraints 
introduced in the companion paper and to the computation of the constraint
algebra.

To put this work in context, we should mention that in a previous
paper \cite{GaGrPu98} we discussed partially the generalization of
Vassiliev invariants to spin networks for the case of trivalent
intersections only. This paper expands and supersedes the discussion
of the previous one. In particular we will discuss several points
which included incorrect interpretations in our previous work. In it,
we mostly concentrated on the definition of the expectation value of a
Wilson loop based on a spin net in a Chern--Simons theory. We
conjectured how one could obtain framing independent invariants
assuming factorizability, and using the notion of primitive Vassiliev
invariants. Further work showed that one cannot really assume
factorizability for intersections of valence four and higher, 
and therefore a small set
\footnote{Specifically, formulas (18) and (30) do not hold.} of the
results of our previous paper do not hold in general, but only for
trivalent intersections. We clarify these issues in the current
paper. In our previous work we also started analyzing the definition
of a diffeomorphism constraint, and, as a warm-up for our current
calculations, the definition of a doubly-densitized Hamiltonian
constraint. However, the final result was dependent on the
regularization and the additional fiducial structures it introduces,
and therefore there was no chance that the operator introduced could
have the correct commutator algebra. Nevertheless, the operator had
some appealing properties, for instance it had the expected
\cite{BrGaPunpb} solution with a cosmological constant (the suitable
generalization of the Kauffman bracket), and annihilated all states
based on trivalent intersections \cite{GaGrPu98}.

In this paper we will extend the definition of Vassiliev invariants to
four-valent spin networks. The four valent case presents considerable
technical challenges, since most literature on spin network invariants
has heavily concentrated on trivalent vertices \cite{Wi89npb,Ma,KaLi}. In
particular the whole issue of constructing framing-independent spin
network invariants starting from Chern--Simons theory has received
virtually no attention (see \cite{GaGrPu97} for first
attempts). Notice that in principle these are the invariants of most
interest for the gravitational case, where wavefunctions are supposed
to be invariant under diffeomorphisms. The crucial need for results
involving four valent intersections stems from the fact that the
volume operator vanishes on states with trivalent intersections. We
will also see, in the companion paper, that the Hamiltonian constraint
vanishes on the framing independent Vassiliev invariants based on
trivalent intersections (we will, however, be able to do calculations
involving the Hamiltonian constraint on related types of wavefunctions
on which the action is non-vanishing even for trivalent spin
networks). We will also show that one can introduce ambient and
regular isotopic invariants in this context, that they are loop
differentiable, and that one can define a suitable diffeomorphism
constraint (infinitesimal generator of diffeomorphisms) on them. We
will also present explicit Feynman-diagrammatic expressions for the
Vassiliev invariants and show that they are finite and well defined
for spin networks.  The latter definition has as a by-product that one
can explicitly compute the loop derivative of these invariants without
recurring to formal manipulations involving the path integral.

In the companion paper we will further elaborate by defining a
single-densitized version of the Hamiltonian constraint in terms of
the loop derivative we define here, we will present several
non-diffeomorphism invariant ``habitats'' where the diffeomorphism
constraint is non-vanishing and we will discuss in detail the
constraint algebra, showing that no anomalies are present. We also
discuss a first application of this construction in the $2+1$
dimensional context.

The organization of this paper is as follows. In the next section
we will briefly introduce the notion of Vassiliev invariants in the
context of ordinary loops. In section III we will discuss in detail
the generalization of Vassiliev invariants to (up to four-valent)
spin-networks and show that one can construct both ambient an regular
isotopic invariants. In section IV we will analyze the loop
differentiability of the invariants and the definition of an
infinitesimal generator of diffeomorphisms on them (diffeomorphism
constraint). We will summarize the results and conclusions
in section V, and also we include an appendix where the action of
the loop derivative over framing independent Vassiliev invariants
is studied.

\section{Vassiliev invariants for loops}

\subsection{Review of perturbative Chern-Simons theory for  ordinary
loops}

We start by recalling several results we will need from ordinary
loops in order to generalize them later to spin networks.  It
is well known that in the context of loops, the expectation value of a
Wilson loop (trace of the holonomy) in a quantum Chern-Simons theory
is related to Vassiliev invariants \cite{BaNa,KaBa}. This expectation
value can be evaluated in various ways. One possible strategy is to
evaluate it perturbatively, taking advantage of the
fact that Chern-Simons theories are perturbatively renormalizable
\cite{GuMaMi}. In fact, it is in the perturbative context where
Vassiliev invariants arise in the most natural way. Let us therefore
consider the expectation value of a Wilson loop $W_A(\gamma) = {\rm
Tr(P}\exp \oint_\gamma dy^a A_a(y))$, where $\gamma$ is a loop in three
dimensions \footnote{The constant $\kappa$ is more commonly written as 
$\kappa=4\pi i/k$, where $k$ is usually taken as an integer for the 
expression to be invariant under large gauge transformations.},
\begin{equation}
<W(\gamma)> = \int DA W_A(\gamma) e^{-{1 \over \kappa} S_{CS}(A)}
\equiv E(\gamma,\kappa,G),
\end{equation}
and,
\begin{equation}
S_{CS}(A) = \int d^3 x {\rm Tr}(A\wedge \partial A +
{2 \over 3} A \wedge A \wedge A),
\end{equation}
This construction defines a regular isotopic knot invariant
\footnote{We are using the following
normalization for the Lie algebra generators
$A_a(x)=A_{ax}^{\alpha}T^{\alpha}$: ${\rm
Tr}[T^{\alpha}T^{\beta}]=\delta^{\alpha\beta}/2$, and $[T^{\alpha},
T^{\beta}]=if^{\alpha\beta\gamma}T^{\gamma}$ with
$f^{\alpha\beta\gamma}$ the structure constants of the group G.} 
$E(\gamma,\kappa,G)$.
This invariant is given by a power series in $\kappa$ and depends
on the gauge group $G$ of the Chern--Simons theory in question. For
$G=SU(2)$ the invariant that appears is (up to a normalization
factor) the Kauffman bracket knot polynomial evaluated for a
particular value of its variable $q= \exp(\kappa)$. One recovers
the power series if one expands the polynomial in terms of
$\kappa$. The coefficients of the power series expansion are linear
combinations of Vassiliev knot invariants\footnote{Originally,
Vassiliev invariants were only considered for the ambient isotopic
case, a straightforward generalization can be done for regular
isotopy (framing dependent invariants).}.

To evaluate the path integral perturbatively, one needs to fix a gauge
and introduce ghosts. It has been established \cite{GuMaMiplb} that if
one considers up to three-point functions, the contributions of the
ghosts produce no radiative corrections. We will therefore ignore them
here, since at the order of perturbation we will work they play no
role. Explicit calculations of the Feynman diagrams 
up to higher orders including ghost
contributions are available in the literature in the context of loops
\cite{GuMaMiplb}. The first relevant Feynman diagram for this theory
corresponds to the two point Green function,
\begin{equation}
<A_a^{\alpha}(x)A_b^{\beta}(y)> =  
\kappa (\delta^{\alpha\beta})( g_{ax\,by})=\kappa
(\,\raisebox{-3mm}{\psfig{file=dot2.eps,height=5mm}}\,)
(\,\raisebox{-3mm}{\psfig{file=prop.eps,height=6mm}}\,).
\end{equation}
The wavy line represents the propagator,
\begin{equation}
g_{ax\,by}\equiv\frac{1}{4\pi}
\epsilon_{abc} {(x-y)^c\over |x-y|^3},
\end{equation}
and the dotted line represents the invariant tensor
$\delta^{\alpha\beta}$. We also have the vertex,
\begin{equation}
{1 \over \kappa} (i f_{\alpha\beta\gamma})(-\epsilon^{abc}\int d^3x)
={1\over \kappa}
(\,\raisebox{-11mm}{\psfig{file=vertdot.eps,height=20mm}}\,)
(\,\raisebox{-10.5mm}{\psfig{file=prueba2.eps,height=20mm}}\,),
\end{equation}
The Lie algebra generators $T^{(R)}_{\alpha}$ (in a given
representation $R$ associated with the weight of the line going from
$A$ to $B$ in the spin net) are represented by the following diagram,
\begin{equation}
(T^{(R)}_{\alpha})^A_B=
\raisebox{-5mm}{\psfig{file=gen2.eps,height=11mm}},
\end{equation}
and from now on we will drop the superscript $(R)$ in the
understanding that it is obvious that the $T$'s are representation
dependent.  In order to evaluate the expectation value of the Wilson
loop, we start by considering the expansion of the path-ordered
exponential,
\begin{eqnarray}
W_A(\gamma) &=&{\rm Tr}\left[ 1 + \oint_\gamma dx^a A_a(x) +
\oint_\gamma dx^a
\int_o^x dy^b A_b(y) A_a(x) \right. \nonumber\\
&&\left. + \oint_\gamma dx^a \int_o^x dy^b \int_o^y
dz^c A_c(z)A_b(y)A_a(x)+\cdots\right].\label{patho}
\end{eqnarray}
If one takes the expectation value of the left hand side, one obtains
in the right hand side a power series in terms of $\kappa$. To zeroth
order one gets,
\begin{equation}
<W(\gamma)>^{(0)} = {\rm dim} R_G,
\end{equation}
which is the dimension of the representation of the group $G$ we are
considering. To first order in $\kappa$ we get,
\begin{equation}
<W(\gamma)>^{(1)} =
\raisebox{-7mm}{\psfig{file=gaupun.eps,height=15mm}}\,\,
\raisebox{-7mm}{\psfig{file=gauond.eps,height=15mm}}\,\,\equiv
r_{11} \alpha^{11}(\gamma)\,,
\end{equation}
where the bold face line represents the Wilson loop $\gamma$, and,
\begin{equation}
\raisebox{-7mm}{\psfig{file=gaupun.eps,height=15mm}}=
{\rm Tr}[T_{\alpha} T_{\beta}] \delta^{\alpha\beta},
\end{equation}
and,
\begin{equation}\label{numero}
\raisebox{-7mm}{\psfig{file=gauond.eps,height=15mm}}\,\, =
\oint_\gamma dx^a \int_o^x dy^b \,g_{ab}(x,y)=
{1 \over 8 \pi} \oint_\gamma dx^a \oint_\gamma dy^b \epsilon_{abc}
{(x-y)^c\over |x-y|^3}\equiv\varphi(\gamma).\label{linknumb}
\end{equation}
The first order contribution factorizes in the product of two
factors\footnote{We use a double index notation because at a given
order there usually appear more than one factor. The first index
denotes the order in the expansion in $\kappa$, the second denotes
which factor within a given order.}: $r_{11}$, which is group and
representation dependent, and $\alpha^{11}(\gamma)$, which depends
only on integrals of the propagator along the Wilson line. They
are called respectively the ``group'' and the ``geometric''
factors. The group factors are given in general by the trace of a
product of Lie algebra generators contracted with the invariant
tensors of the group. They can be thought of as the insertion at
points in the Wilson line of the Lie algebra generators, and the
evaluation of the resulting net for a flat connection. This
diagrammatic representation of the group factors is usually called a
``chord diagram''. The geometric factor $\alpha^{11}(\gamma)$
corresponds to the self-linking number $\varphi(\gamma)$ of the
loop. The factorization in terms of a group and a geometric factor is a
general property of the perturbative expansion of $<W(\gamma)>$.

For second order in $\kappa$ we have the following contributions,
\begin{equation}\label{2orden}
<W(\gamma)>^{(2)} =
\raisebox{-7mm}{\psfig{file=cir2parp.eps,height=15mm}}\,\,
\raisebox{-7mm}{\psfig{file=parond.eps,height=15mm}}
+\raisebox{-7mm}{\psfig{file=cruzpun.eps,height=15mm}}\,\,
\raisebox{-7mm}{\psfig{file=circ2cru.eps,height=15mm}}
+\raisebox{-7mm}{\psfig{file=cir2trip.eps,height=15mm}}\,\,
\raisebox{-7mm}{\psfig{file=circ2tri.eps,height=15mm}}\,,
\end{equation}
with,
\begin{eqnarray}
\raisebox{-7mm}{\psfig{file=cir2parp.eps,height=15mm}} &=&
{\rm Tr}[T_{\alpha} T^{\alpha} T_{\beta} T^{\beta}]\equiv r_{21},\\
\raisebox{-7mm}{\psfig{file=cruzpun.eps,height=15mm}} &=&
{\rm Tr}[T_{\alpha} T_{\beta} T^{\alpha} T^{\beta}],\\
\raisebox{-7mm}{\psfig{file=cir2trip.eps,height=15mm}} &=&
{\rm Tr}[T_{\alpha}  T_{\beta} T_{\gamma}] if^{\alpha\beta\gamma}
\equiv r_{22}.\label{3pgfloop}
\end{eqnarray}
The analytic expression of the geometric factors are obtained
directly from the diagrammatic representation. For example,
\begin{equation}
\raisebox{-7mm}{\psfig{file=circ2cru.eps,height=15mm}}={1\over 4}
[\oint_\gamma dx^a \int_o^x du^b \int_o^u
dy^c \int_o^y dw^d g_{ac}(x,y)g_{bd}(u,w)+ \mbox{\footnotesize{c.p.
}}],
\end{equation}
where $c.p.$ means to consider terms obtained by cyclic permutations
of the first term in the indices $(ax)(bu)(cy)(dw)$.

The three chord diagrams included in (\ref{2orden}) are
not all independent. They are related by the so called ``STU''
relations \cite{BaNa},
\begin{equation}
\raisebox{-7mm}{\psfig{file=cruzpun.eps,height=15mm}}=
\raisebox{-7mm}{\psfig{file=cir2parp.eps,height=15mm}}+
\raisebox{-7mm}{\psfig{file=cir2trip.eps,height=15mm}}\,,
\label{stu}
\end{equation}
which stems from the commutation relation of the group generators.
These relations allow to eliminate all overlapping diagram in terms
of a combination of non-overlapping diagrams. Proceeding in this
way one can define a canonical basis for the group factors
\cite{AlLaPe}. In terms of this basis, the second order
contribution can be written in the following way,
\begin{equation}
<W(\gamma)>^{(2)}=
r_{21} {1 \over 2} [\alpha^{11}(\gamma)]^2+r_{22}
\alpha^{22}(\gamma),\label{invloops}
\end{equation}
where
\begin{equation}
\alpha^{22}(\gamma)\equiv
\raisebox{-7mm}{\psfig{file=circ2tri.eps,height=15mm}}\,\, +
\raisebox{-7mm}{\psfig{file=circ2cru.eps,height=15mm}}\,.\label{17}
\end{equation}
In the above result two things should be noticed. The first is
that, via the application of the STU relations, the geometric
factors associated with the independent group factors will include
in general a contribution coming from the overlapping diagrams. In
particular, the combination of geometric factors associated with
$r_{21}$ factorizes in a product of first order contributions,
\begin{equation}
\alpha^{21}(\gamma)\equiv
\raisebox{-7mm}{\psfig{file=parond.eps,height=15mm}}\,\, +
\raisebox{-7mm}{\psfig{file=circ2cru.eps,height=15mm}}={1\over
2}\raisebox{-7mm}{\psfig{file=gauond.eps,height=15mm}}^2
\label{factloop}
\end{equation}
Moreover, applying recoupling \cite{DeRo} identities
\footnote{For general results about the
application of recoupling theory to the factorization property of
group factors, see equations (\ref{fierz}) and (\ref{recoup})
below.} one can show that
\begin{equation}
r_{21}=\frac{1}{<W(\gamma)>^{(0)}}[r_{11}]^2,
\end{equation}
so the first term in (\ref{invloops}) equals the square of the
first order contribution $[W(\gamma)>^{(1)}]^2/2<W(\gamma)>^{(0)}$.

The second observation is that, as $r_{ij}$ depend on the group and
representation chosen, this implies that the geometric factors
$\alpha^{ij}(\gamma)$ embody the invariance under diffeomorphisms
individually. That is, each $\alpha^{ij}(\gamma)$ is an independent
knot invariant. We can also reach some conclusions concerning the
nature of the constructed invariants. In principle, the integrals
involving wavy lines have to be regularized to avoid having
propagators with two of their ends evaluated at the same point
\cite{GuMaMi}. In the context of knot theory this regularization
corresponds to a framing of the knot.  A possible framing consists,
for instance, in prescribing a copy of the loop infinitesimally
displaced away from the original one along a vector field on the
manifold, and evaluating the various integrals along the separate
loops. This procedure is not needed for all invariants. For instance,
if we in detail look at $\alpha^{22}(\gamma)$ we notice that in the
evaluation of the integrals involved, the two ends of each propagator
can never coincide.  In the first term, one of the ends is always at
the vertex, in the second term the ordering of the integrals prevents
the ends from coinciding, if they do, one of the accompanying
integrals is evaluated along a loop of zero length and vanishes.  We
refer to this case as saying that there are no ``collapsible''
propagators. The invariants that do not require the framing procedure
are true diffeomorphism invariants of loops, and are called in the
knot theory language {\em ambient} isotopic invariants. The invariants
that require of the framing procedure are invariants of ``ribbons''
(instead of loops), and are called {\em regular} isotopic invariants.

The explicit expressions for the higher order contributions
are progressively more complicated as $\kappa$ increases.
However, at any order one can apply essentially the same procedure
we discussed above to isolate the independent group factors, and
the geometric factors associated with them. The
expectation value can therefore be generically written as,
\begin{equation}
<W(\gamma)> = \sum_{i=0}^\infty \sum_{j=1}^{d_i} r_{ij}
\alpha^{ij}(\gamma) \kappa^i \,,
\end{equation}
where $d_i$ is the number of independent invariants at order
$\kappa^i$. The invariant associated with $<W(\gamma)>$ is a
regular isotopic invariant, that is, it is framing dependent. One
can extract the framing dependence in terms of an overall phase
factor \cite{Wi89cmp}, and be left with a framing independent
invariant $J(\gamma)$ defined by,
\begin{equation}
\exp\left(-\kappa{<W(\gamma)>^{(1)} \over <W(\gamma)>^{(0)}}\right)
<W(\gamma)> = J(\gamma).\label{expon}
\end{equation}
For the case of $SU(2)$, $J(\gamma)$ is the Jones polynomial for a
particular value of its variable $q=\exp(\kappa)$ and expanded as a
power series in terms of $\kappa$. We will return to this issue in
terms of spin networks later.

\subsection{Vassiliev invariants}

Let us now show that the invariants that appear at order $\kappa^i$
in the perturbative evaluation of the expectation value of the
Wilson loop are Vassiliev invariants of order $i$. 
This is a well known
result in the case of ordinary loops (see for instance
\cite{KaBa}). To do this, we recall the definition of Vassiliev
invariants \cite{Va}. Given a knot invariant $v^n(\gamma)$, one starts
by introducing a ``Vassiliev'' intersection (which we will illustrate
with a black square in diagrams), defined as,
\begin{equation}
v^n(\raisebox{-7mm}{\psfig{file=vasint.eps,height=15mm}}) \equiv
v^n(\raisebox{-7mm}{\psfig{file=upcross.eps,height=15mm}}) -
v^n(\raisebox{-7mm}{\psfig{file=dwcross.eps,height=15mm}}). 
\end{equation}
The invariant $v^n(\gamma)$ is called {\em Vassiliev invariant of
order $n$}, if and only if when evaluated with any knot with $n+1$
Vassiliev intersections, it vanishes. One can immediately see that it
also vanishes for all knots containing more than $n+1$ intersections. 
This is the reason the invariants are sometimes called ``of finite
type''. This is also related to the observation that the Vassiliev
intersection can be viewed as the action of a differential operator.
We will see later on that this abstract differential operation ends
up remarkably being link to the loop derivative.

To see that the invariants we constructed via perturbation theory are
Vassiliev invariants, we need certain properties of the path integral
that can be derived by variational techniques
\cite{Sm88,CoGuMaMi,BrGaPunpb,KaBa,GaPucmp}. For the sake of brevity we will
not repeat the whole argument here. We will just recall that one can
convert an upper in an under crossing in the expectation value by
considering the action of the loop derivative (which appends a small
loop). If one does this, one derives the following identity,
\begin{equation}
<W(\raisebox{-7mm}{\psfig{file=vasint.eps,height=15mm}})> \equiv
<W(\raisebox{-7mm}{\psfig{file=upcross.eps,height=15mm}})> -
<W(\raisebox{-7mm}{\psfig{file=dwcross.eps,height=15mm}})> =
\kappa <h(\gamma_1) T_{\alpha} h(\gamma_2) T^{\alpha}> +O(\kappa^2)
\end{equation}
The meaning of this equation is as follows: any crossing divides
a loop $\gamma$ into two subloops, $\gamma_1$ and $\gamma_2$, according
the connectivity of the loop. The invariant
of the right is constructed replacing the holonomy along the whole
loop by a holonomy divided into the subloops with insertions of the
Lie algebra generators. If one now considers $n$ Vassiliev
intersections, one can repeat the above construction for each
intersection, i.e.,
\begin{equation}
<W(\raisebox{-7mm}{\psfig{file=vasint.eps,height=15mm}},\stackrel{n
\,times}{\ldots},\raisebox{-7mm}{\psfig{file=vasint.eps,height=15mm}})>
= \kappa^n <h(\gamma_1) T_{\alpha_1} h(\gamma_2)
T_{\alpha_2} \cdots h(\gamma_k)
T^{\alpha_1}h(\gamma_{k+1})\ldots>+O(\kappa^{n+1}),
\end{equation}
where now $\gamma_1, \gamma_2, \ldots$ are the arcs that join one
crossing with the following. The generators are inserted in each
crossing starting from the origin. Notice that each crossing is
traversed twice, and that the two generators associated with a given
crossing are contracted between themselves. We see that the left
hand side of the above equation has vanishing coefficients up to
order $n$, that is, the coefficients of the expansion up to order
$n$ behave like Vassiliev invariants of order equal or less than
$n$. A similar behavior is observed for the case of spin networks.

\section{Vassiliev invariants for spin networks}
\subsection{Diagrammatic notation for the expectation value of a
Wilson net}

A spin network \cite{RoSm95} is constructed considering a graph
$\Gamma$ (with multivalent intersections) embedded in a three
dimensional manifold. The graph is composed by a set of edges
$\{e_j\},\,(j=1,\ldots,N(\Gamma))$, which are analytic oriented
open paths that intersect only in the initial and final points. The
intersection point of three (or more) edges defines a tri (or
multivalent) vertex $v$. We denote $N(\Gamma)$ the number of edges,
and $V(\Gamma)$ the number of vertices of the graph. Associated
with each edge $e_i$ in the graph is a representation labelled by
$J_i$ of an element of the gauge group $G$. For instance, given a
connection, one can construct these elements by considering the
holonomy along the given edge in a given representation. The group
elements are ``intertwined'' into gauge invariant objects through
contraction at the vertices of the network with the invariant
tensors of the group. Given a graph $\Gamma$ and an assignment
$\vec{J}=(J_1,\ldots,J_{N(\Gamma)})$ of representations to each edge
plus a set of intertwiners $\vec{I}=(I_1, \ldots,I_{V(\Gamma)})$,
one can construct starting from a connection an object we call
``Wilson net'', $W_A (s)\equiv W_A (\Gamma,\vec{I},\vec{J})$, in
the following way,
\begin{equation}\label{wnet}
W_A (s)\equiv \left[ \prod_v {I_v}^{C_1\ldots C_{n_v}}_{B_1\ldots
B_{m_v}}\right]\left[\prod_e
U(e)^{B_e}_{C_e}\right].
\end{equation}
We will use the compact notation $s$ denoting the graph,
intertwiners and spin weights together. In the above expression,
all the matrix indices of the holonomies $U(e)$ are contracted with
the indices of the intertwiners $I_v$ in a way which depends on
the connectivity of the net. The edges of the net are oriented
according to the following conventions,
\begin{equation}
U(e)^{B_e}_{C_e}= \raisebox{-11mm}{\psfig{file=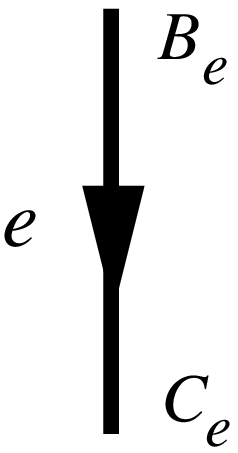,height=22mm}},
\end{equation}
and,
\begin{equation}
{I_v}^{C_1\ldots C_{n_v}}_{B_1\ldots B_{m_v}}=
\raisebox{-11mm}{\psfig{file=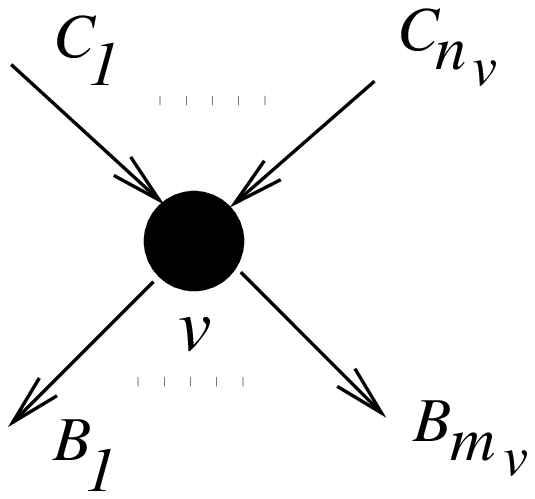,height=22mm}}.
\end{equation}
The invariant tensor ${I_v}^{C_1\ldots C_{n_v}}_{B_1\ldots
B_{m_v}}$ can be expressed in a basis of $n_v+m_v-2$ Clebsch-Gordan
intertwiners\footnote{In general, a $n+m$-valent intersection can
be thought of as the limit where the intermediate lines are of 
vanishing length,  of $n+m-2$ trivalent intersections
connected by additional links, in a choice of basis such that the
intermediate links are assigned a definite spin value.}. This
construction is feasible for any group, but we will 
concentrate in the following sections 
on $SU(2)$ and the representations will be labelled by
integers. A more elaborate labeling is needed for other groups.
Spin networks were originally introduced in the $SU(2)$ context by
Penrose \cite{Roger}.

We now wish to compute the expectation value of a Wilson net in a
Chern-Simons theory,
\begin{equation}
<W(s)> = \int DA W_A(s) e^{-{1 \over \kappa} S_{CS}(A)}
\equiv E(s,\kappa). \label{intspin}
\end{equation}
This integral has been studied by Witten and Martin \cite{Ma} and
Kauffman and Lins \cite{KaLi}, establishing the crossing relations
(skein relations) and recoupling relations for networks including up to
trivalent intersections. Their calculations were based on the use of
monodromies and the tangle group, respectively. For trivalent vertices
we have also shown in a previous paper \cite{GaGrPu97} that one can
extract the framing factor and construct ambient isotopic
invariants.

Here we would like to study if analogous results can be found in the
case of four-valent intersections, and also perform perturbative
studies of the path integral with the aim of constructing explicitly
ambient invariants.  We therefore proceed as in the case of loops, to
consider order by order the perturbative evaluation of expression
(\ref{intspin}).  The general form of the perturbative expansion can
be easily deduced from the definition (\ref{wnet}). We first introduce
a convenient diagrammatic notation. The holonomy associated with a
generic edge $e$ is written in the following form, introducing a 
diagrammatic notation for the explicit expression of the path ordered
exponential in terms of the connection,
\begin{equation}
U(e)^{B_e}_{C_e}=
\raisebox{-11mm}{\psfig{file=hol.eps,height=20mm}}=\delta^{B_e}_{C_e}
+ \sum_{n=1}^{\infty}
\raisebox{-7mm}{\psfig{file=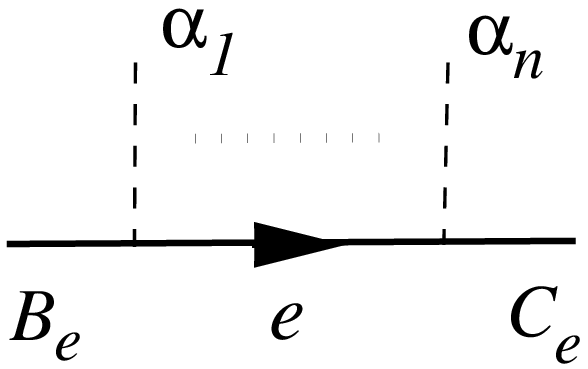,height=14mm}}\,
\raisebox{-5.8mm}{\psfig{file=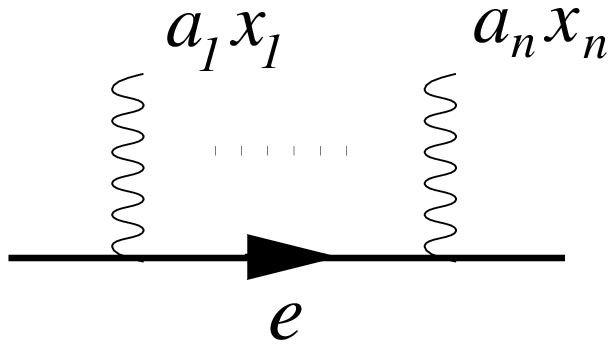,height=14mm}}\,
A^{\alpha_1}_{a_1 x_1}\ldots A^{\alpha_n}_{a_n x_n},
\label{holdia}
\end{equation}
where,
\begin{equation}
\raisebox{-6mm}{\psfig{file=dibujo1.eps,height=14mm}}\,
= (T^{(J_e)}_{\alpha_1}\ldots
T^{(J_e)}_{\alpha_n})^{B_e}_{C_e},
\end{equation}
and,
\begin{equation}
\raisebox{-5mm}{\psfig{file=dibujo2.eps,height=14mm}}\,
=\int_e dy_n^{a_n} \int_o^{y_n} dy_{n-1}^{a_{n-1}}\ldots
\int_o^{y_2} dy_1^{a_1} \delta(x_n-y_n)\ldots\delta(x_1-y_1).
\end{equation}
For the readers familiar with the extended loop formalism 
(\cite{GaPubook} chapter 2), the above multitangent corresponds to 
$X(e)^{a_1x_1\cdots a_n x_n}$ in such a notation. In the above 
expressions, we have adopted the usual ``generalized Einstein 
convention'' of extended loops, in which a summation is implied by 
any repeated index $a_i$ and an integral over the three manifold
is implied by any repeated index $x_i$.

Some general results about the factorization of the geometric
factors (see equation (\ref{factloop})) can be derived from the
following property of these fields,
\begin{equation}
\sum_{P(1\ldots m,m+1\ldots n)}
\raisebox{-5mm}{\psfig{file=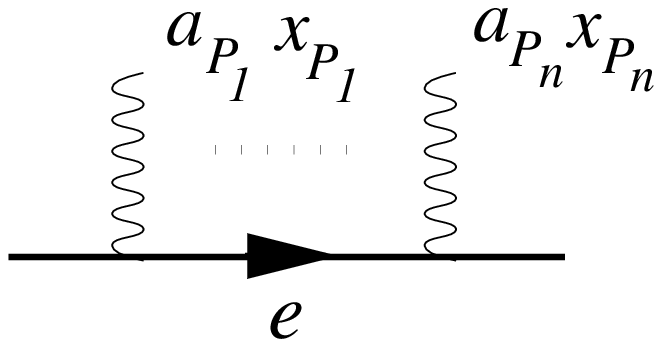,height=14mm}}\,
=\raisebox{-5mm}{\psfig{file=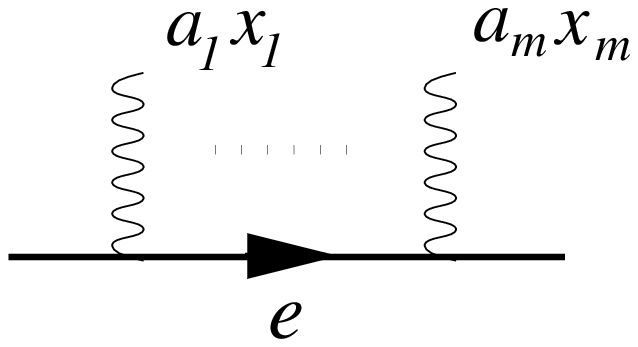,height=14mm}}\times
\raisebox{-5mm}{\psfig{file=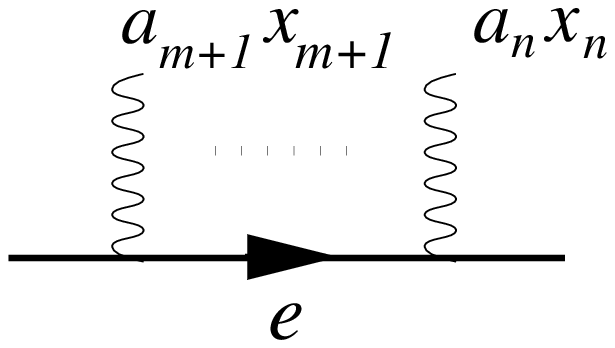,height=14mm}},
\label{ac}
\end{equation}
where $P(1\ldots m,m+1\ldots n)$ is any permutation of the indices
which preserves the ordering of the subsets $(1\ldots m)$ and
$(m+1\ldots n)$ between themselves. This equation express the
``algebraic constraint'' of the multitangent fields, first studied
in \cite{DiGaGr}. Introducing (\ref{holdia}) in (\ref{wnet}), we
get the following diagrammatic expression for the expectation value
of the Wilson net,
\begin{eqnarray}
<W_A (s)>&=&
\sum_{n_1=0}^{\infty}\ldots
\sum_{n_{N(\Gamma)}=0}^{\infty}
W_{A={\rm flat}}\left(
\,\raisebox{-6mm}{\psfig{file=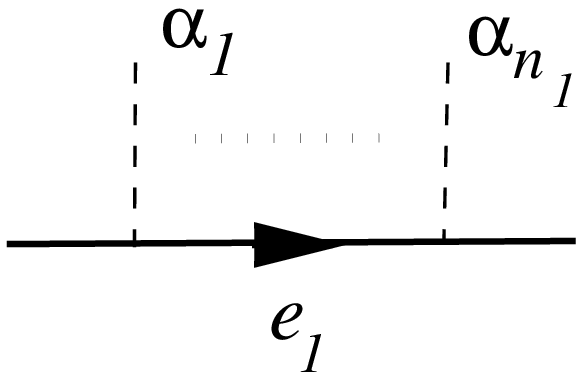,height=13mm}}\,
\ldots \right) \left[
\raisebox{-5.5mm}{\psfig{file=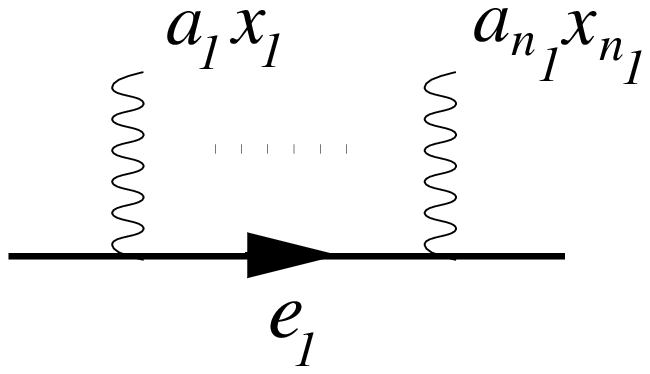,height=13mm}}\,
\ldots\right]\times\nonumber\\
&&\times 
<(A^{\alpha_1}_{a_1 x_1}\ldots A^{\alpha_{n_1}}_{a_{n_1}
x_{n_1}}) \ldots >,
\label{diawn}
\end{eqnarray}
Notice that in the above equation one takes a sum for each edge of
the net. In general, $W_{A={\rm flat}}(s)$ defines the ``chromatic
evaluation'' of the spin network $s$. This quantity is evaluated
replacing in the functional integral all the holonomies by identity
matrices, $U(e)^{B_e}_{A_e}\rightarrow \delta^{B_e}_{A_e}$ for all
$e$. The resulting expression can be computed using recoupling
theory. Then, in (\ref{diawn}) $W_{A={\rm flat}}(
\raisebox{-0.3mm}{\psfig{file=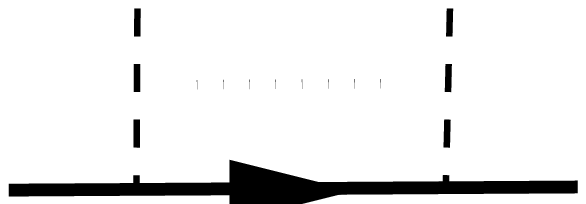,height=3mm}}\, \ldots)$ has
the following meaning: take the chromatic evaluation of the spin
network replacing the holonomy $U(e_i)$ by a product of Lie algebra
generators $T_{\alpha_1}\ldots T_{\alpha_{n_i}}$ in the
representation $J_i$ if $n_i\neq 0$, or by the identity matrix if
$n_i=0$.

At this point it is that the convenience of the notation we introduced
becomes clear, we see in equation (\ref{diawn}) that the expression
for the expectation value of the Wilson net becomes really compact,
involving a chromatic evaluation of a diagram times another diagram
contracted with the propagators. This is quite analogous to what
happened in the case of ordinary Wilson loops. To generate from
(\ref{diawn}) the perturbative series one has to introduce the Feynman
rules to express the Green functions $<(A^{\alpha_1}_{a_1 x_1}\ldots
A^{\alpha_{n_1}}_{a_{n_1} x_{n_1}}) \ldots >$ in terms of wavy lines
(the propagator and the vertex)\footnote{As in the case of loops, we
will ignore the ghosts. It one wishes to include them, one has to
incorporate in the analysis the ghost's propagator and the ghost's
vertex.} and dotted lines (the invariant tensors of the group). We
immediately see that a mechanism similar to that observed in the case
of loops operates here: the contraction of the invariant tensors with
$W_{A={\rm flat}}(
\raisebox{-0.3mm}{\psfig{file=dibujo11.eps,height=3mm}} \ldots)$
generates dotted diagrams (the group factors), whereas the contraction
of the two and three point propagators with the multitangent fields
generates wavy diagrams (the geometric factors). We schematically
represent this procedure in the following way,
\begin{equation}
W_{A={\rm flat}}\left(
\raisebox{-5mm}{\psfig{file=dibujo10.eps,height=13mm}}\ldots \right)
\cdot[\delta^{\alpha_1\alpha_{n_1}}\ldots]\equiv
r\left(\raisebox{-5mm}{\psfig{file=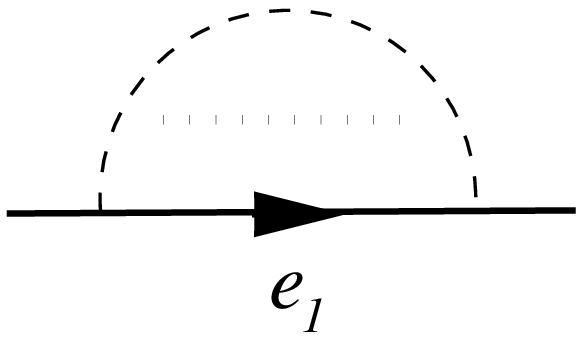,height=13mm}}\ldots
\right)\equiv r(\mbox{\footnotesize{dotted diagram}}),
\end{equation}
and,
\begin{equation}
\left[\raisebox{-5mm}{\psfig{file=dibujo20.eps,height=13mm}}\ldots\right]
\cdot(g_{a_1 x_1 a_{n_1} x_{n_1}}\ldots)
\equiv\left(
\raisebox{-5mm}{\psfig{file=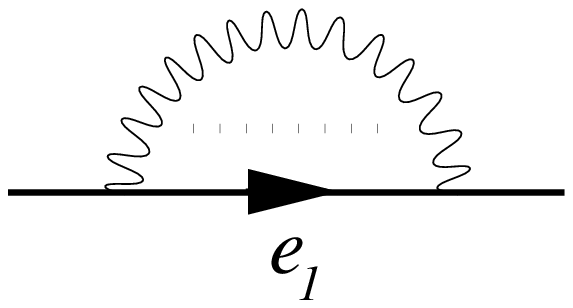,height=12mm}}\ldots \right)
\equiv (\mbox{\footnotesize{wavy diagram}}),
\end{equation}
These relations define the group and geometric factors for spin
networks. Notice that in the above examples we have chosen a
specific way to connect the indices $\alpha_1, \alpha_{n_1}$ and
$a_1 x_1, a_{n_1} x_{n_1}$. In general, the application of the
Feynman rules produce, for a given number of indices, a
superposition of diagrams with different connectivities of the
dotted (or wavy) lines. Therefore, we can schematically write
(ignoring the ghosts as we discussed before),

\begin{equation}
<W(s)>=\sum_{m=0}^{\infty}\sum_{\{\mbox{\scriptsize{diagrams}}\}}
r[(\mbox{\scriptsize{dotted diagrams}})^{(m)}]
(\mbox{\scriptsize{wavy diagrams}})^{(m)}\, \kappa^m,
\label{exwn}
\end{equation}
The index $m$ gives the order of the perturbative expansion, and
for a fixed $m$ we sum over all the diagrams that contributes to
this order. Each term factorizes in the product of a dotted diagram
(the group factor) and a wavy diagram (the geometric factor), both
having the same connectivity (i.e., in both diagrams the connection
of the dotted and wavy lines are identical).

In principle, the group factors can be evaluated using recoupling
theory. For $SU(2)$ one uses the following version of the Fierz
identity (the standard identity between the Lie algebra generators
and the Clebsch--Gordan coefficients in the fundamental
representation),
\begin{equation}
T^{(J)}_{\alpha} T^{(K)}_{\alpha} =
\raisebox{-5mm}{\psfig{file=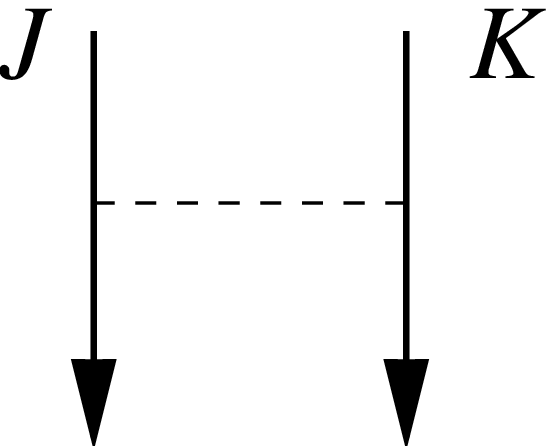,height=12mm}} =
(-1)^{2(J+K)+1}\Lambda_{JK}
\raisebox{-5mm}{\psfig{file=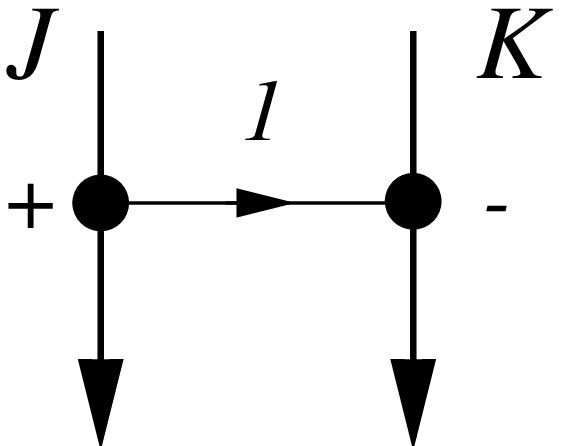,height=12mm}},\label{fierz}
\end{equation}
where $\Lambda_{JK}=\Lambda_{J}\Lambda_{K}$ and
$\Lambda_{J}=\sqrt{J(J+1)(2J+1)}$. The Fierz identity allows
to rewrite any dotted diagram in terms of the chromatic evaluation of
a new spin network where the dotted lines are substituted by edges
of spin one. In general, the group factors satisfy some identities
which are characteristic of spin networks and which play an
important role at the time to identify the topological invariants
associated with the perturbative expansion. To introduce these
identities, let us consider a generic dotted diagram and let us
isolate from it a single dotted line connecting two edges $e$ and
$e'$. We draw this single line enclosed into a circle. We also draw
one of the vertices of the selected edges, for example, the origin
of $e'$, which we assume to be four-valent. The dotted lines
outside the circle remain fixed as we move the generator starting
from $e'$ all around the vertex. Then it is easy to prove that,
\begin{equation}
r\left(\raisebox{-6mm}{\psfig{file=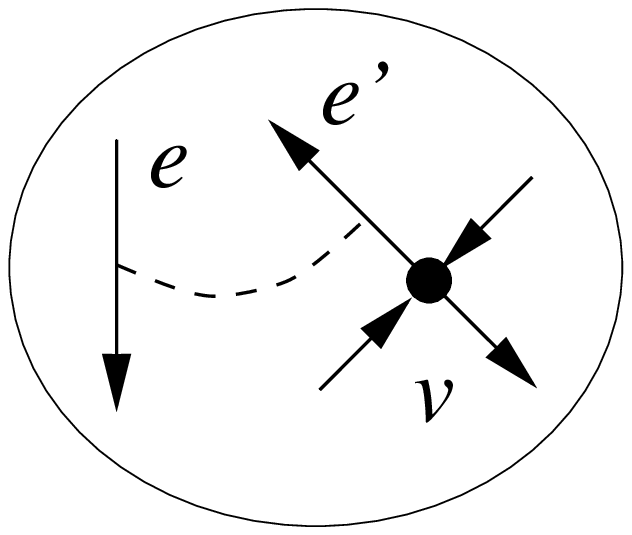,height=15mm}}\right)-
r\left(\raisebox{-6mm}{\psfig{file=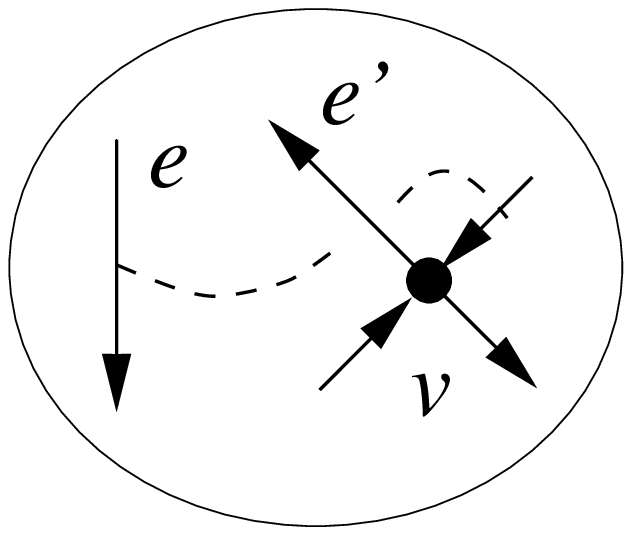,height=15mm}}\right)+
r\left(\raisebox{-6mm}{\psfig{file=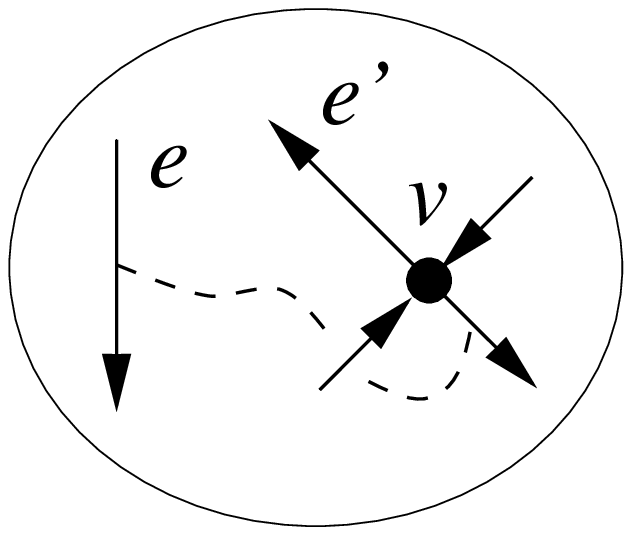,height=15mm}}\right)
-r\left(\raisebox{-6mm}{\psfig{file=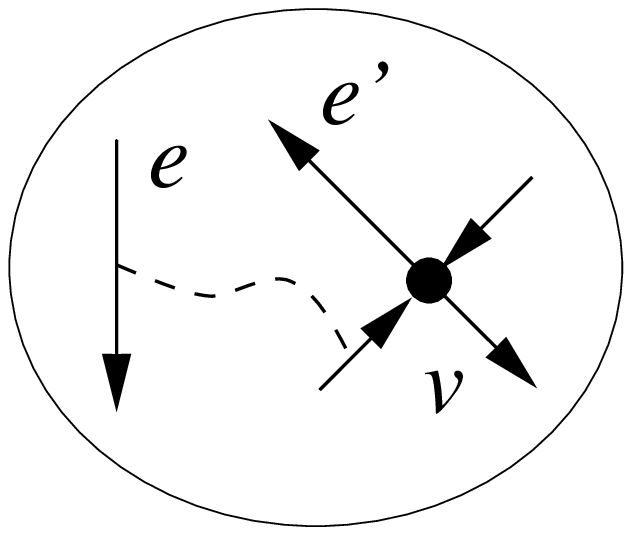,height=15mm}}\right)
\equiv 0
\label{gauss}
\end{equation}
where $r$ is any of the group factors we have considered (we omit
indices since the expression is valid at any order).  For a given
diagram, one can perform these movements with all the dotted lines of
the graphs producing a linear system of equations for the group
factors (of course, not all the possible movements generate
independent equations). This set of identities is valid for any group,
and they are rooted in the gauge invariance property of the Wilson
net. For this reason, we call these identities the Gauss' law for spin
networks.

Another set of identities, called ``STU relations'' arise from the generator algebra,
\begin{equation}
\raisebox{-2mm}{\psfig{file=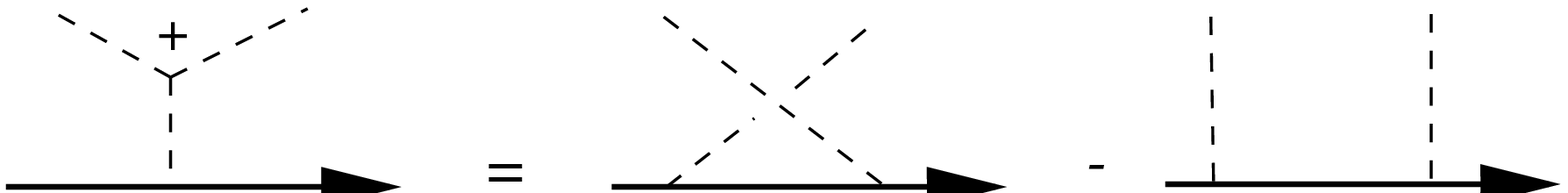,height=10mm}}
\end{equation}

The STU identities, together with Gauss' law, form a set of linear
identities that can be solved order by order in the perturbative
expansion, allowing to find all group factors in terms of a reduced
set of ``independent factors''.

The zeroth order contribution of (\ref{exwn}) is given by the
chromatic evaluation of the net, which we usually denote in the
form,
\begin{equation}
<W(s)>^{(0)} = W_{A={\rm flat}}(s)\equiv E(s,0)
\end{equation}
In the following subsections we are going to study the detailed
structure of the group and geometric factors for the first and
second order in $\kappa$.

\subsection{The first order contribution}

Considering equation (\ref{exwn}) to first order in $\kappa$ we have,
\begin{equation}
<W(s)>^{(1)} = 
\sum_i \sum_{j\leq i}
r\left(\raisebox{-4mm}{\psfig{file=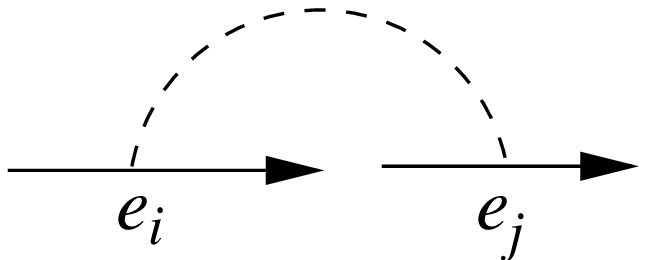,height=9mm}}\right)
\raisebox{-4mm}{\psfig{file=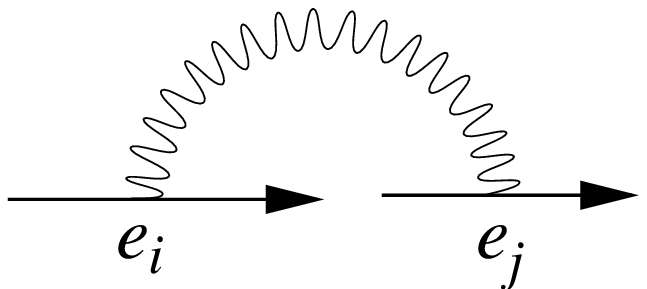,height=10mm}}
\equiv
\sum_i \sum_{j\leq i}
r_{ij} \varphi_{ij},
\label{25}
\end{equation}
where by $i$ and $j$ we denote any pair of edges of the spin network
(they can also be the same edge). We usually use the compact notation
$r_{ij}$ for the first order group factors, and $\varphi_{ij}$ for a
wavy diagram connecting the edges $e_i$ and $e_j$.

We notice a difference with the case of loops. In that case there
exists only one group factor, $r^{11}_G$, at first order. Here we have
a number of first order group factors equals to
$N(\Gamma)(N(\Gamma)+1)/2$.  These quantities are not all independent
due to the Gauss constraints (\ref{gauss}). The Gauss identities
generate for a given graph a linear system of equations which allows
to express some of the $r_{ij}$ in terms of a small subset, which we
call the set of ``independent'' group factors (IGF). The determination
of the IGF allows to construct the topological invariants associated
with the spin network. As in the case of loops, the linear
combinations of geometric factors that multiply each group factor are
(regular) knot invariants.

It is clear that the IGF set is not unique (we can
choose different free parameters to express the solutions of the
Gauss identities).

Let $\mbox{IGF}=\{z_i\}$, $i=1\ldots d(s)$, be the free
parameters with respect to which the linear system of Gauss
equations is solved. The number $d(s)$ of parameters depend
on the particular spin network being considered. Then, (\ref{25})
can be rewritten in the following way,
\begin{equation}
<W(s)>^{(1)} = \sum_{i=1}^{D(s)} z_i {\cal I}_i(\varphi)
\end{equation}
where the ${\cal I}(\varphi)$'s represent knot invariants formed by 
the linear combination of
geometric factors associated with each independent parameter.

\begin{figure}
\centerline{\psfig{figure=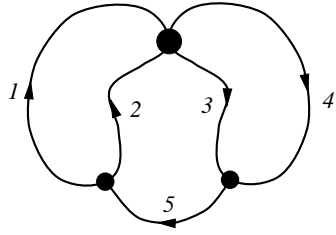,height=30mm}}
\caption{A spin net with a four-valent and a trivalent vertices.}
\label{ejemplo}
\end{figure}

Let us illustrate the procedure with an example.
For the net shown in figure \ref{ejemplo}, $d(s)=6$ and we choose
$\mbox{IGF}=\{r_{12},r_{13},r_{14},r_{23},r_{24},r_{34}\}.$
Then one gets
\begin{eqnarray}
<W(\raisebox{-5mm}{\psfig{file=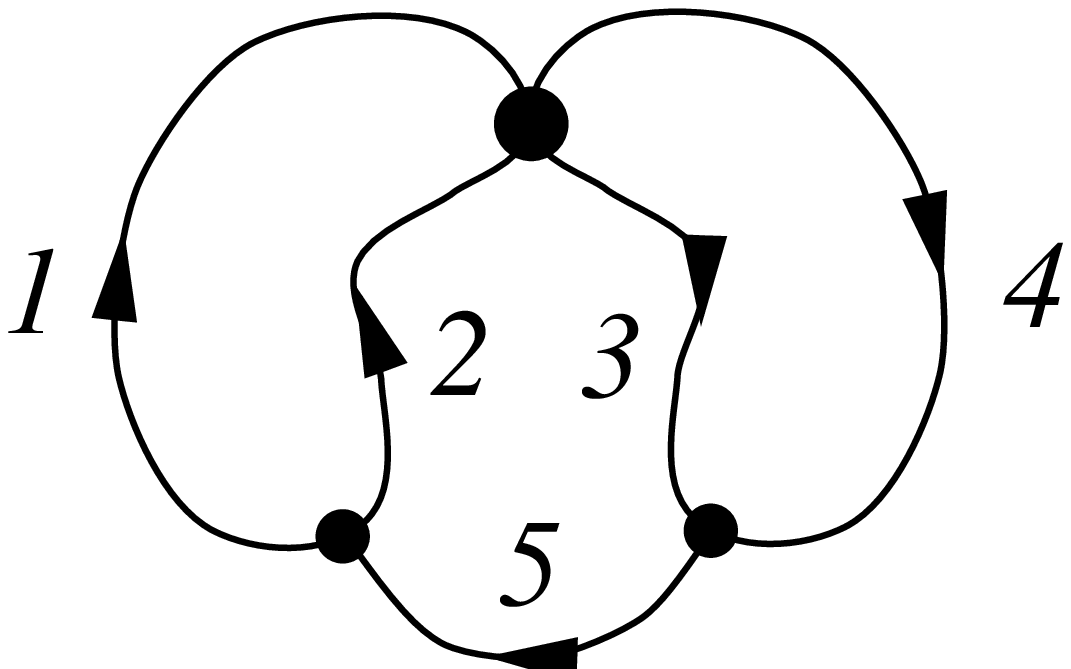,height=10mm}})>^{(1)}&=&
-r_{12}(\varphi_{11}+\varphi_{22}-\varphi_{12})
-r_{34}(\varphi_{33}+\varphi_{44}-\varphi_{34})\nonumber\\
&&+\sum_{i=1}^{2}\sum_{j=3}^{4} r_{ij}(
\varphi_{ii}+\varphi_{jj}-\varphi_{55}+
\varphi_{i5}+\varphi_{j5}+\varphi_{ij}).
\end{eqnarray}
The invariants have a simple geometric interpretation in terms of
linking numbers of loops constructed with the edges of the graph.
It is immediate to see that each of the above linear superposition of
geometric factors combine to give the self-linking number of a
closed path $\gamma$ with support in the net,
\begin{equation}
<W(\raisebox{-5mm}{\psfig{file=ejred.eps,height=10mm}})>^{(1)}=
-r_{12}\varphi(\gamma_{12})-r_{34}\varphi(\gamma_{34})
+\sum_{i=1}^{2}\sum_{j=3}^{4} r_{ij}
\varphi(\gamma_{ij5}),
\label{28}
\end{equation}
with\footnote{Overbars denote edges traversed in opposite
directions, and the $\circ$ operation is the usual composition law
of open paths.},
\begin{eqnarray}
\gamma_{ij}&\equiv&e_i\circ\bar{e}_j\label{gammaij}\\
\gamma_{ij5}&\equiv&e_i\circ e_j\circ e_5 \label{gammaij5}
\end{eqnarray}
For any other IGF, the invariants are written as a combination of
the above self-linking numbers. Notice that in this example all the
first order invariants are framing dependent.

\subsection{The second order contribution}

Considering the contribution to  second order in $\kappa$ in 
(\ref{exwn}) we have,
\begin{eqnarray}\label{secondorder}
&&<W(s)>^{(2)} = \sum_i \sum_j^i \sum_k^j
r\left(\raisebox{-5mm}{\psfig{file=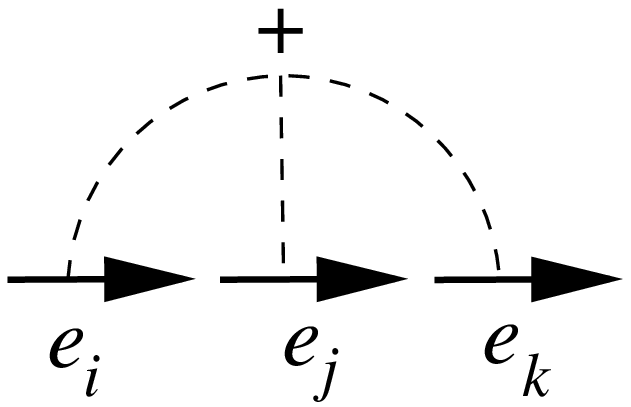,height=12mm}}\right)\left[
\raisebox{-5mm}{\psfig{file=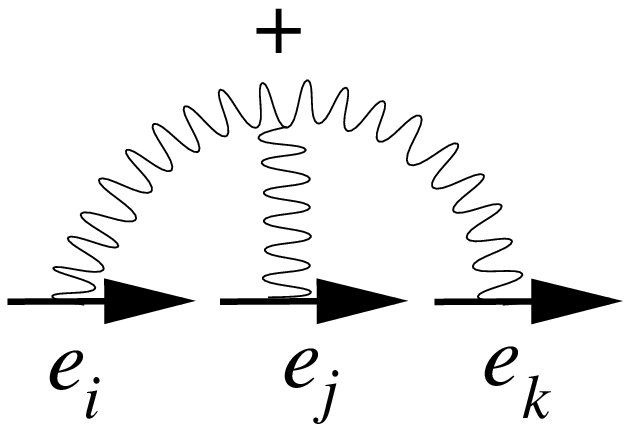,height=12mm}}\right]+\\
&&\sum_i \sum_j^i \sum_k^j \sum_l^k
\left\{
r\left(\raisebox{-7mm}{\psfig{file=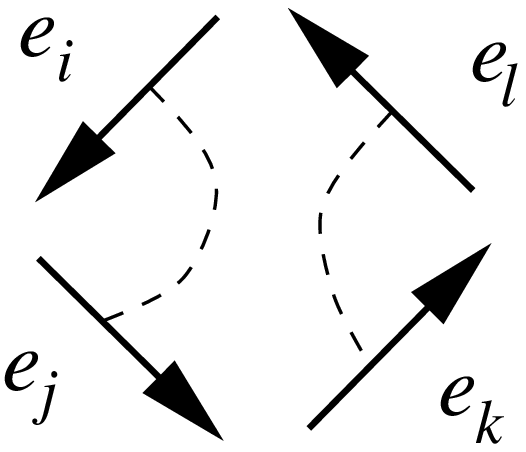,height=14mm}}\right)
\left[\raisebox{-7mm}{\psfig{file=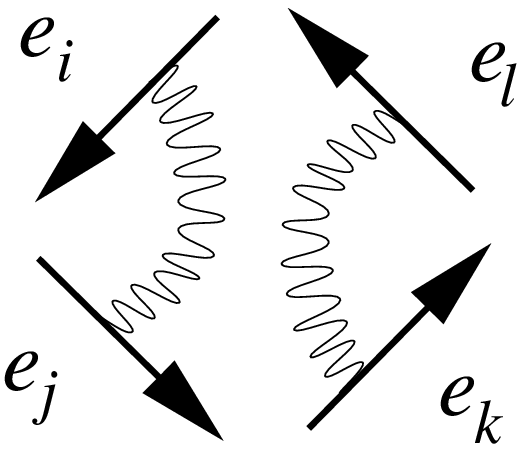,height=14mm}}\right]+
r\left(\raisebox{-7mm}{\psfig{file=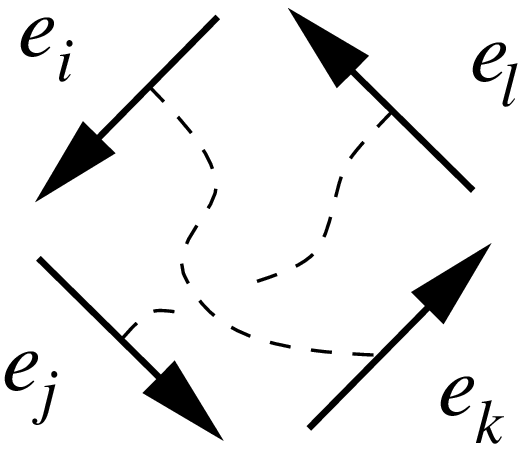,height=14mm}}\right)
\left[\raisebox{-7mm}{\psfig{file=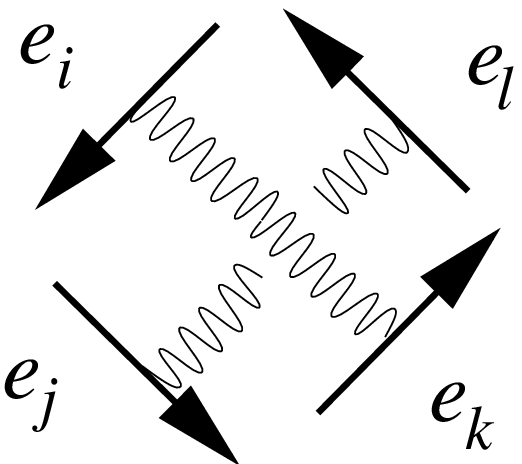,height=14mm}}\right]\right.
\nonumber\\
&&\left.+
r\left(\raisebox{-7mm}{\psfig{file=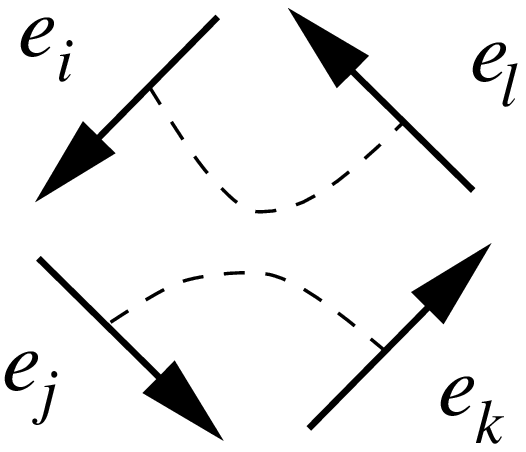,height=14mm}}\right)
\left[\raisebox{-7mm}{\psfig{file=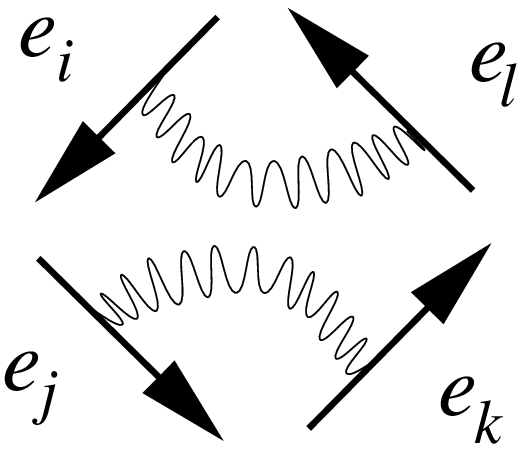,height=14mm}}\right]\right\}.
\nonumber\end{eqnarray}

Guided by the case of ordinary loops, we would like to study if it is possible to extract a pre-factor that is dependent on framing, therefore isolating a framing-independent knot invariant in this expression. In the case of ordinary
loops the exponential of the first order contribution was the prefactor that
allowed to isolate the Jones polynomial. Here we would like to check if this
is possible for spin nets up to second order in the expansion, for four
valent intersections. We will see that, as we independently proved in our
previous paper, it is possible to extract the prefactor for trivalent vertices,
but it is not possible for four and higher valent intersections. This was
already anticipated in the work of Labastida, who noted that it does not
hold for multiloops, which are a particular case of spin networks.

Using the Gauss and STU identities and the algebraic constraint we discussed
before, and observing that,
\begin{equation}
r\left(\raisebox{-6mm}{\psfig{file=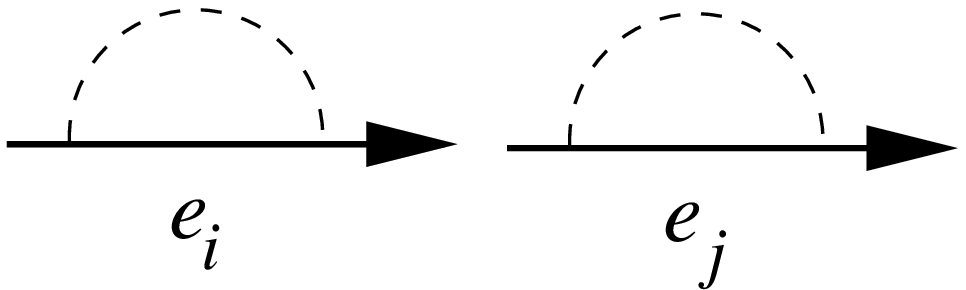,height=12mm}}\right)=
{1 \over E(s,0)} 
r\left(\raisebox{-6mm}{\psfig{file=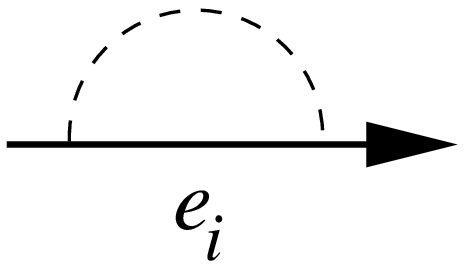,height=12mm}}\right)
r\left(\raisebox{-6mm}{\psfig{file=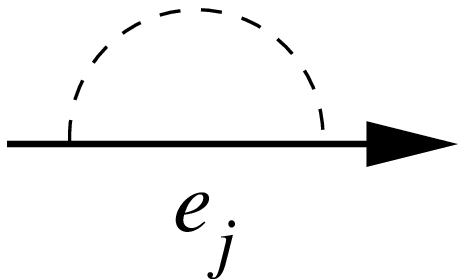,height=12mm}}\right),
\label{recoup}
\end{equation}

and introducing the ``shifted'' four-point group factors
$\tilde{r}$, defined in the
following way,
\begin{equation}
\tilde{r}(
\raisebox{-6mm}{\psfig{file=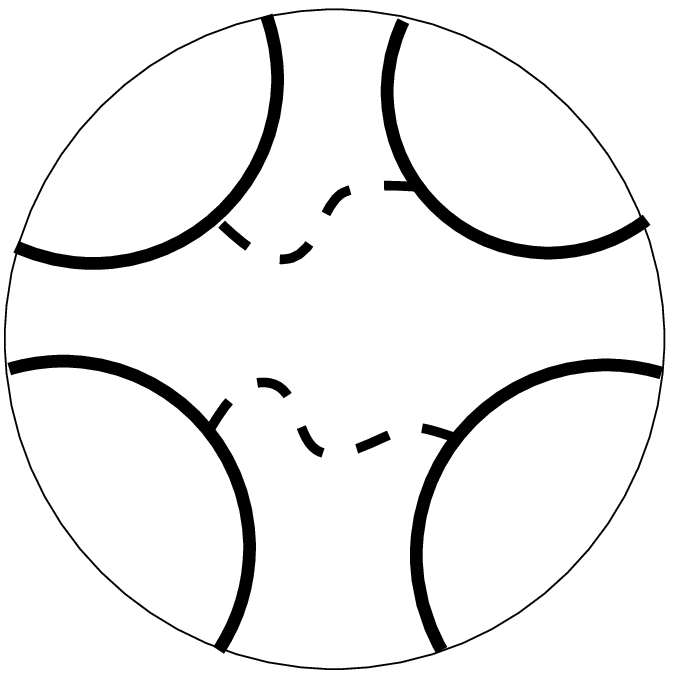,height=13mm}})=
r(\raisebox{-6mm}{\psfig{file=rtilde.eps,height=13mm}})-
\frac{1}{E(s,0)}r(
\raisebox{-6mm}{\psfig{file=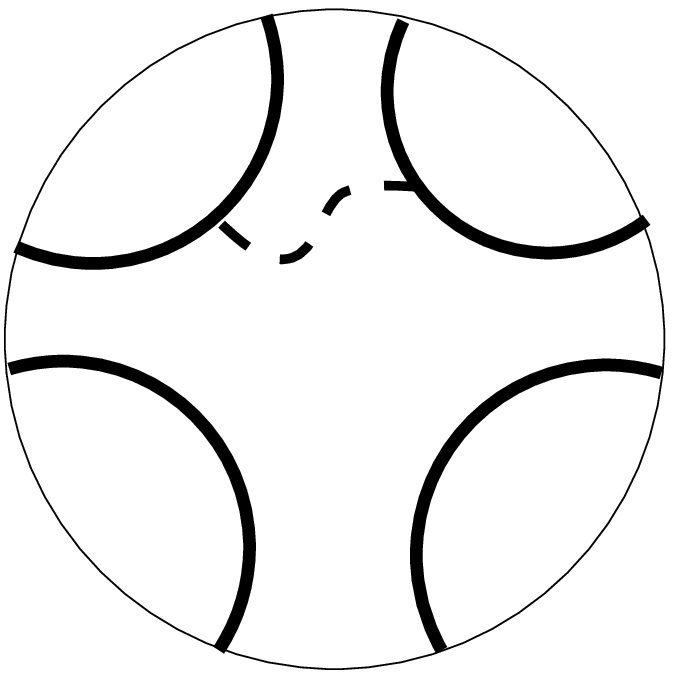,height=13mm}})
r(\raisebox{-6mm}{\psfig{file=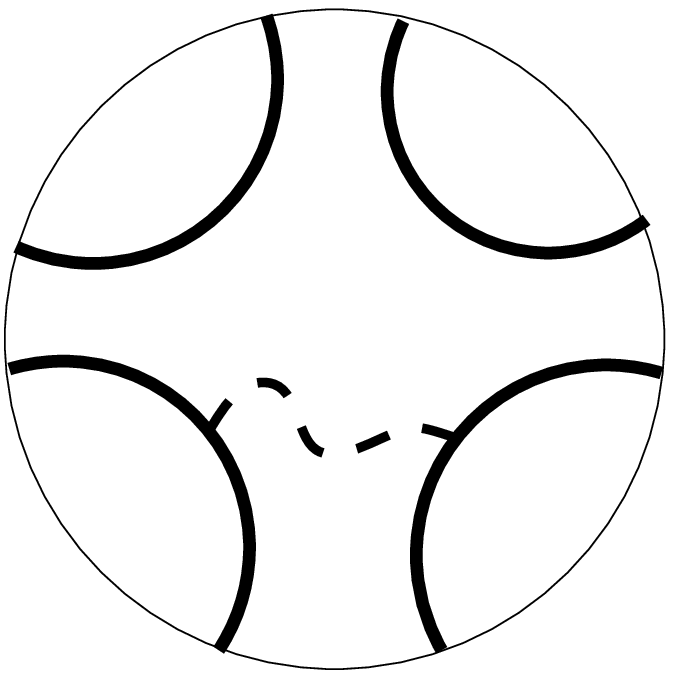,height=13mm}}),\label{rtilde}
\end{equation}
it is found that,
\begin{eqnarray}
<W(s)>^{(2)} &=& \frac{1}{2E(s,0)}[<W(s)>^{(1)}]^2 \nonumber\\
&&+
\sum_i \sum_j^{i-1} \left\{
\tilde{r}\left(
\raisebox{-5mm}{\psfig{file=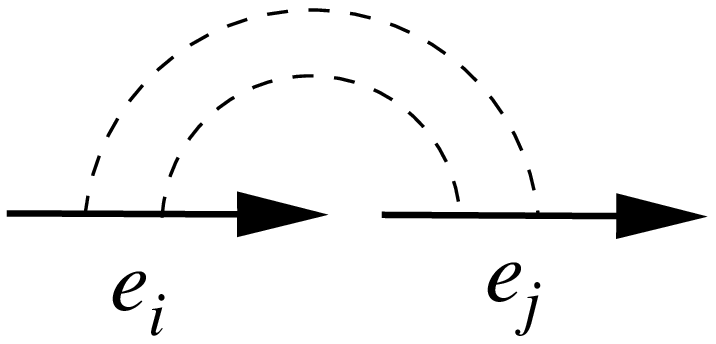,height=12mm}} \right)
\frac{1}{2}[\varphi_{ij}]^2 
+r\left(
\raisebox{-6.5mm}{\psfig{file=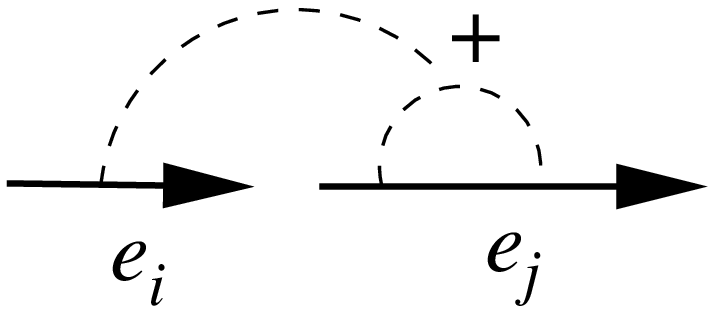,height=12mm}}
\right)\rho_{ij}\right\}
\nonumber\\
&&+ \sum_i\sum_j^{i-1}\sum_k^{j-1} \left\{\tilde{r}\left(
\raisebox{-5.5mm}{\psfig{file=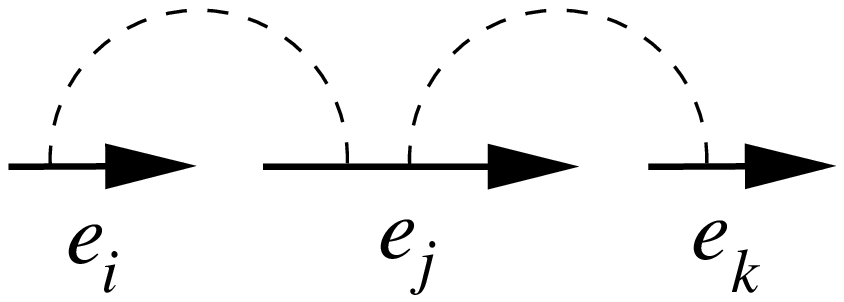,height=12mm}} \right)
\varphi_{ij}\varphi_{jk}+\hbox{cyclic permutations in i,j,k}\right\}
\nonumber\\
&&+\sum_i
r\left(\raisebox{-6.5mm}{\psfig{file=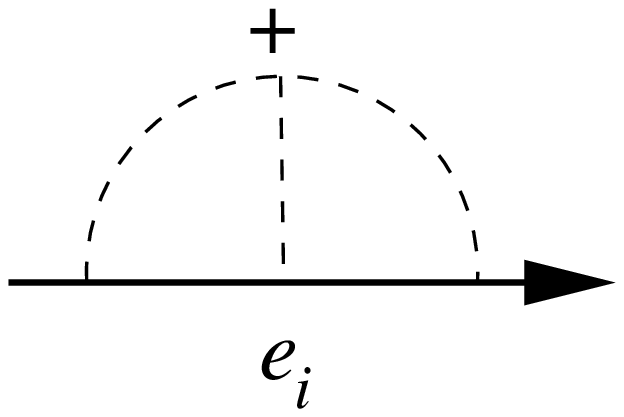,height=12mm}}\right)
\rho_{i}
+
\sum_i\sum_j^{i-1}\sum_k^{j-1}r\left(
\raisebox{-6.5mm}{\psfig{file=3patasd.eps,height=12mm}}\right)
\rho_{ijk},
\label{so}
\end{eqnarray}

The quantities $\varphi_{ij}$ are the first order
geometric factors linking the edges $e_i$ and $e_j$, and,
\begin{equation}
\rho_{i}\equiv
\raisebox{-5mm}{\psfig{file=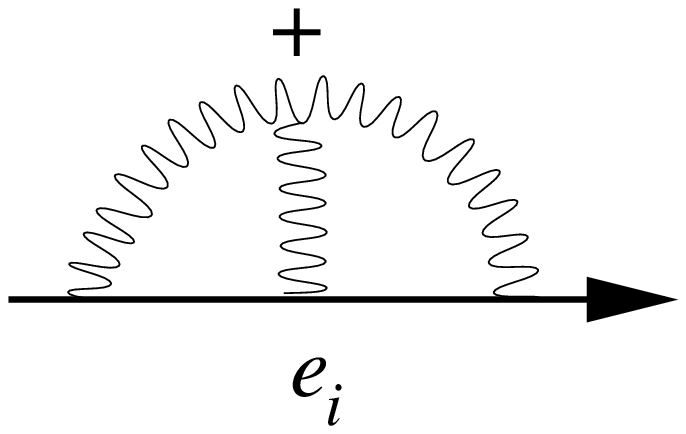,height=12mm}}
+\raisebox{-6mm}{\psfig{file=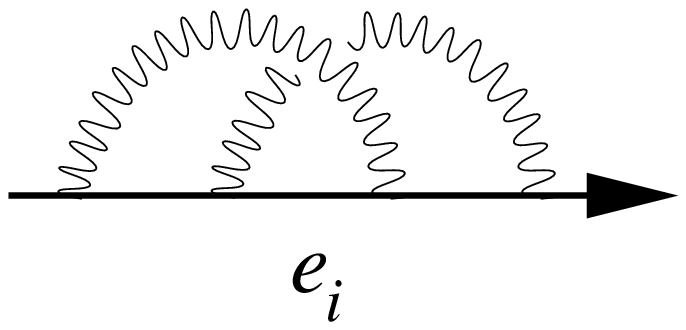,height=12mm}},
\end{equation}
\begin{eqnarray}
\rho_{ij}&\equiv&
\raisebox{-5mm}{\psfig{file=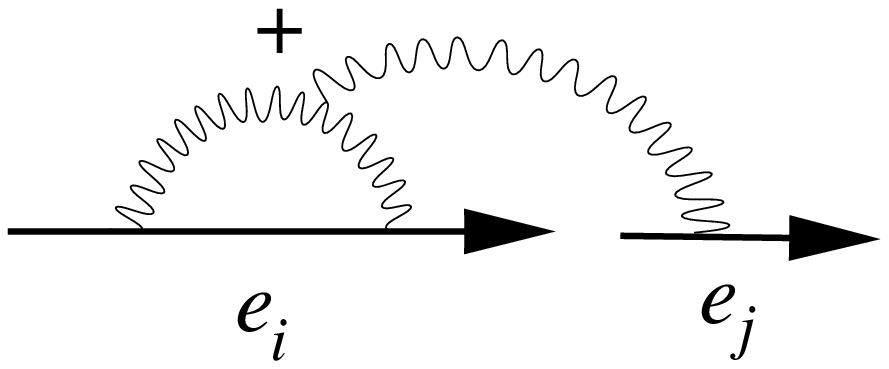,height=12mm}}
+\raisebox{-5mm}{\psfig{file=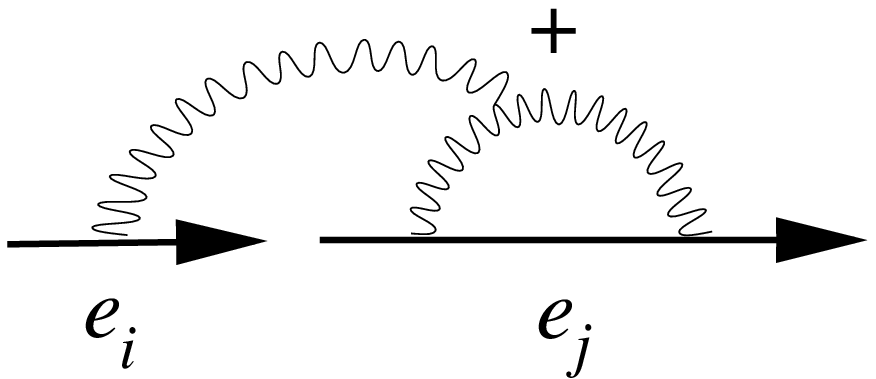,height=12mm}}
\\
&&+\raisebox{-5mm}{\psfig{file=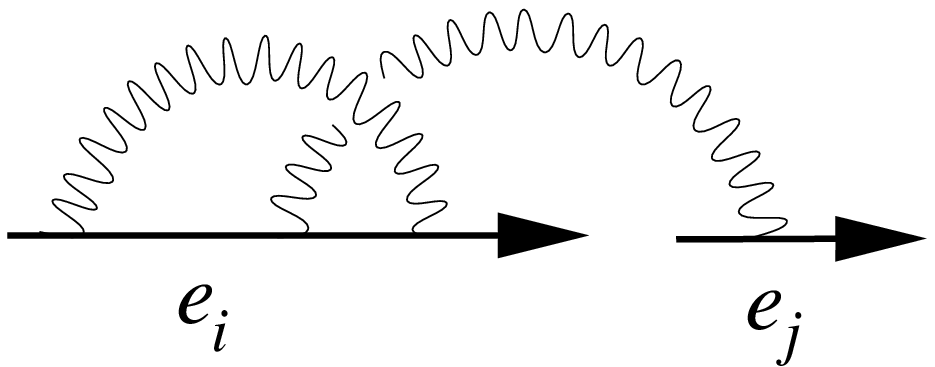,height=12mm}}
+\raisebox{-5mm}{\psfig{file=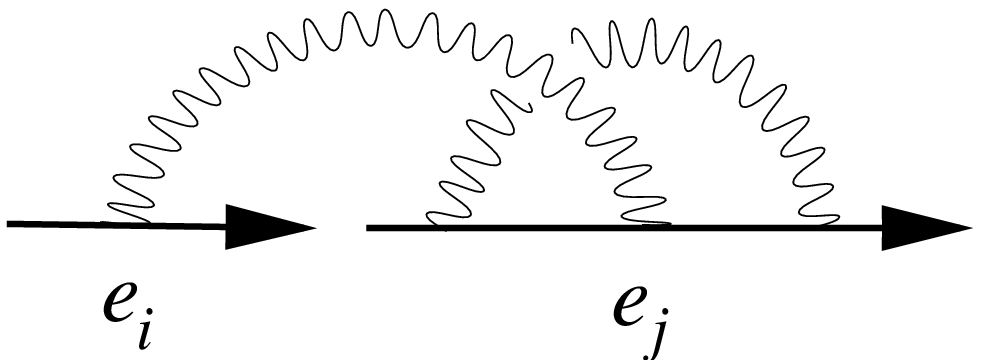,height=12mm}}
+\raisebox{-5mm}{\psfig{file=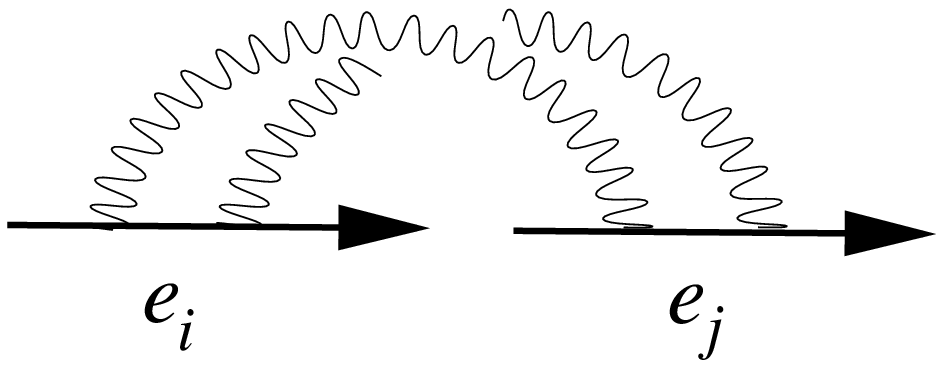,height=12mm}},
\nonumber
\end{eqnarray}
and,
\begin{equation}
\rho_{ijk}\equiv
\raisebox{-5mm}{\psfig{file=3patasw.eps,height=12mm}}
+\raisebox{-5mm}{\psfig{file=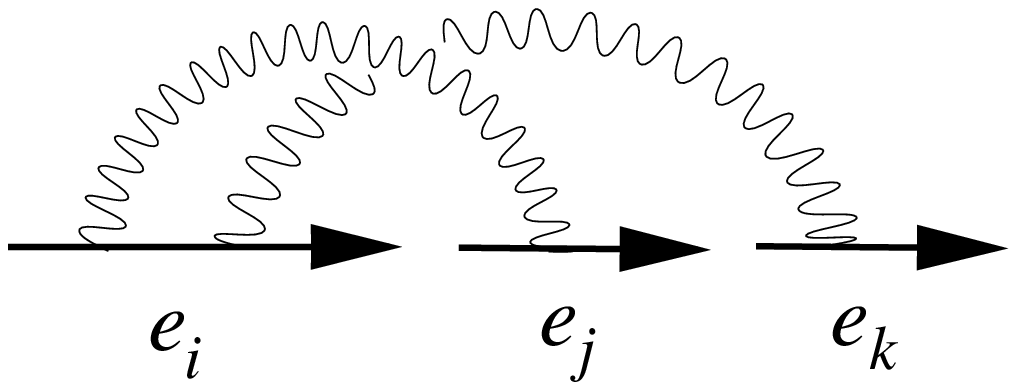,height=12mm}}
+\raisebox{-5mm}{\psfig{file=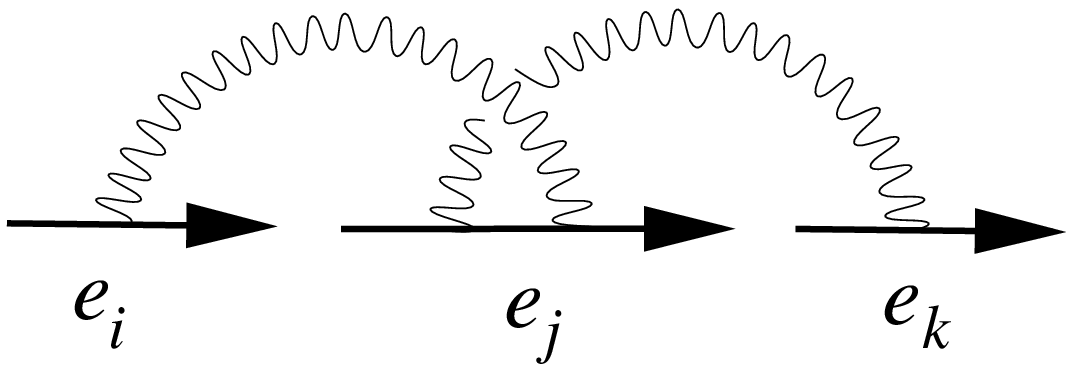,height=12mm}}
+\raisebox{-5mm}{\psfig{file=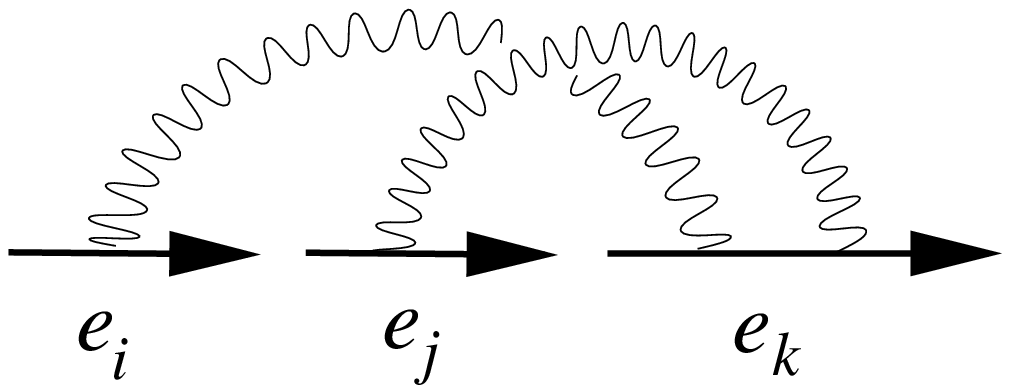,height=12mm}}.
\end{equation}

Looking at equation (\ref{so}) we see that we have not achieved the
desired result. If we look at the first term in the right hand side of
this equation we identify the expansion to second order of the
exponential of the first order correction, exactly like in the case of
loops. However, there remain framing dependent contributions in the
other terms, therefore one will not be able to extract the framing
dependence in an overall prefactor, as happened in the case of
loops. The additional framing dependence arises for instance in the
terms involving 
$\varphi_{ij}$ or $\rho_{ijk}$ when the edges $i,j,k$
share a common vertex. 

A remarkable exception happens in the case of trivalent
intersections. In this case the group factors $r$ that appear in the
above mentioned terms vanish identically. This can be seen
straightforwardly applying recoupling at the vertex. In the case of
trivalent vertices this yields only a prefactor and the two terms involved
in the definition of the $\tilde{r}$, equation (\ref{rtilde})
vanish. In the case of higher order vertices, the application of
recoupling leads to a sum of terms and therefore the two terms in
(\ref{rtilde}) are different. This fact will also appear at the root
of our proof that the loop derivative annihilates trivalent framing
independent invariants, as we will discuss later on.

\subsection{Topological invariants for spin networks}

The analysis of the higher order contributions can be performed
using the same techniques we have developed at first and second
order. The results can be summarized as follows.  To any order in the
perturbative expansion, there exist a number of independent group
factors for a given group $G$, $z_a^{(n)}(\vec{J},\vec{I},G)$, based
on solving the Gauss and STU relations. In terms of these the
perturbative expansion can be written as,
\begin{equation}
<W>^{n}=\sum_{a=1}^{d^n(s)} z_a^{(n)}(\vec{J},\vec{I},G) {\cal
I}_a^{(n)}(\Gamma) 
\end{equation}
where the ${\cal I}_a^{(n)}$ are (regular or ambient) isotopic
invariants of the embedding of the spin network $\Gamma$, 
and $d^n(s)$  is the number of independent group factors at
$n-th$ order for a given spin network $s$. This is
a difference with the case of ordinary loops, where the expressions
are given once and for all loops. This is related to the fact that
the spin nets are a much more general object. 

The regular or ambient isotopy properties of
the invariants can be analyzed systematically, and, to any order in
$\kappa$, the invariants are grouped in two categories: framing
independent and framing dependent. The framing dependent part
cannot be exponentiated like in the case of loops. The general
result can be written in the following way,
\begin{equation}
<W(s)> = \sum_{i=0}^\infty
\sum_{a=1}^{d^{\rm fd}_i(s)}z_{ia}^{\mbox{\scriptsize{fd}}}
(\vec{J},\vec{I},G)
\,{\cal I}_{ia}^{\mbox{\scriptsize{fd}}}(\Gamma)\, \kappa^i +
\sum_{i=0}^\infty \sum_{b=1}^{d^{\rm fi}_i(s)}
z_{ib}^{\mbox{\scriptsize{fi}}}(\vec{J},\vec{I},G)
\,{\cal I}_{ib}^{\mbox{\scriptsize{fi}}}(\Gamma)\, \kappa^i.
\label{decomp}
\end{equation}

In spite to this
dependence with the particular spin network considered, the
invariants have the following property: given an embedding
$\Gamma$, let $\Gamma'\in \Gamma$ be a subgraph obtained by
removing some of the edges of $\Gamma$. Then, the invariants ${\cal
I}(\Gamma')$ are obtained from ${\cal I}(\Gamma)$ putting zero all
the wavy diagrams with legs attached to any of the suppressed
edges. 

The specific expression of the invariants ${\cal I}_{fd}$ and ${\cal
I}_{fi}$ that appear in expression (\ref{decomp}) and their quantity
$d_n^{fi}(s)$, depend on the choice of independent invariants picked
to solve the constraints among the group factors. We will give a
procedure that generates, for a given spin network, a basis of
independent invariants that are framing-independent. 

A general invariant of $n-th$ order is in general 
given by a linear combination,
\begin{equation}
{\cal I}^{(n)}(\Gamma)= \sum_{a=1}^{d^n(s)} c_a {\cal I}_a^{(n)}(\Gamma),
\end{equation}
where the ${\cal I}_a$'s have the following expression,
\begin{equation}
{\cal I}_a^{(n)}(\Gamma)=\sum_i D_{ia}^{\rm fi} \alpha^{(n){\rm
fi}}_i(\Gamma)
+\sum_i D_{ia}^{\rm fd} \alpha^{(n){\rm fd}}_i(\Gamma),
\end{equation} 
where the  $D_{ia}^{\rm fd, fi}$ are well defined numbers that depend
on the connectivity of the spin net and the $\alpha$'s are products of
multitangents contracted with the geometric portion of the Feynman
diagrams (propagators and vertices) and can either be framing
dependent or independent. We can therefore conclude that an invariant
will be framing independent if the choice of coefficient $c_a$'s is
such that,
\begin{equation}
\sum_{a=1}^{d^n(s)} c_a D_{ja}^{\rm fd}=0.
\end{equation}
One now solves this system of equations, which will lead to expressing
some of the $c_a$'s in terms of a set of independent $c_a$'s. For each
of these independent $c_a$'s one has an independent invariant.

Let us consider a concrete example of this procedure. We will carry
out the computation for the spin network of the figure (\ref{ejemplo}).
The invariants that appear at second order in the calculation have the
generic form,
\begin{equation}
{\cal I}^{(2)}(\Gamma)= c_1 \rho(\gamma_{12}) + c_2 \rho(\gamma_{31})
+\sum_{i=1}^2 \sum_{j=i}^4 c_{ij} \rho(\gamma_{ij5}),
\end{equation}
where $\gamma_{ij}$ are the subloops of the spin network we defined in
equations (\ref{gammaij},\ref{gammaij5}). The quantities $\rho$ are
closely related with the second coefficient of the Alexander-Conway
polynomial \cite{GuMaMi}. These were the expressions one directly got
when one performed the calculation in the case of loops. In the case
of spin networks one gets combinations of these invariants evaluated
in sub-loops that form the spin network. The expression is manifestly
framing independent since the coefficients of the Alexander-Conway
polynomial are framing-independent. We see that (at least up to this
order in the perturbative expansion) the use of spin
networks does not introduce new topological invariants, but just
combinations of invariants evaluated in the various subloops that form
the spin network.

Finally, we show that the invariants we have obtained are the
natural generalization to the spin network context of Vassiliev
invariants. To see this, consider an arbitrary net, and construct
the Vassiliev intersection. Following the same type of argument as
we did in the case of ordinary loops one immediately obtains, using
variational techniques,
\begin{eqnarray}
\left<W\left(\raisebox{-7mm}
{\psfig{file=svasint.eps,height=15mm}}\,\,\right)\right> 
\label{vasivasi}
&\equiv&
\left<W\left(\raisebox{-7mm}
{\psfig{file=supcross.eps,height=15mm}}\,\,\right)\right> -
\left<W\left(\raisebox{-7mm}
{\psfig{file=sdwcross.eps,height=15mm}}\,\,\right)\right> \nonumber=\\
&&= \kappa
< \ldots h^{(J_j)}(e_{1j}) T^{(J_j)}_{\alpha} h^{(J_j)}(e_{2j})\ldots
h^{(J_k)}(e_{3k}) T^{(J_k)}_{\alpha}h^{(J_k)}(e_{4k})\ldots>
\end{eqnarray}
where the indices in parenthesis refer to the representation in which
one is considering the holonomies and the Lie algebra basis
elements. We therefore reach the same conclusion as in the case of
loops: an invariant with $n$ Vassiliev crossings is of order
$\kappa^n$ meaning that the first $n-1$ coefficients in the series
expansion vanish for spin networks with $n$ crossings. Therefore, they
are Vassiliev invariants of order up to $n-1$. 

\section{The diffeomorphism constraint}

\subsection{Loop differentiability}

We now have succeeded in setting a suitable ``arena'' where we can
start discussing the constraints of quantum gravity. The arena will be
the space of Vassiliev invariants, and in order to implement the
constraints as quantum operators we will take advantage of the fact
that these invariants are loop differentiable.  This is a result we
introduced in a previous paper \cite{GaGrPu98}, we only recall the
appropriate formula here. The result appears simply by applying the
definition of loop derivative \cite{GaPubook} to the path integral
defining the expectation value of the Wilson loop. One acts with
the loop derivative on the Wilson net in the path integral. One
obtains as a result a Wilson net with an $F_{ab}$ inserted at the
point where the loop derivative acted. One then re-expresses the
$F_{ab}$ in terms of a functional derivative of the exponential of
the Chern--Simons form with respect to the connection. This
functional derivative is integrated by parts to act on the Wilson
loop again. The result is the expectation value of a Wilson net
with the insertion of two Lie algebra generators,
\begin{eqnarray}
&&\Delta_{ab}(\pi_o^x) E(\raisebox{-5mm}{\psfig{file=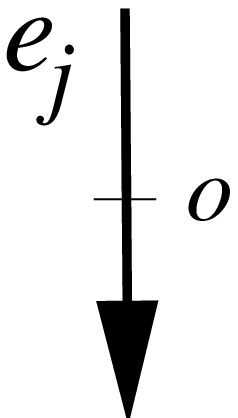,height=11mm}},
\kappa) = -2 \kappa \sum_{k}
\epsilon_{abc} \int_{e_k} dy^c \delta^3(x-y) \times
\label{tata}\\
&&< \cdots U^{(J_j)}(e_{v_j}^o) U^{(J_j)}(\pi_o^x) T^{(J_j)}_{\alpha}
U^{(J_j)}({\pi^{-1}}{}_o^x) U^{(J_j)}(e_o^{v'_j})\cdots U^{(J_k)}
(e_{v_k}^y)T^{(J_k)}_{\alpha} U^{(J_k)}(e_y^{v'_k}) \cdots>,\nonumber
\end{eqnarray}
where $v_j$ and $v'_j$ are the start and end points of edge
$e_j$. The loop derivative depends on a path $\pi_o^x$ going from
the basepoint $o$ to a point $x$, and this operator acts on the
space of spin networks with a fixed basepoint $o\in\Gamma$. The links
involved in the formula are shown in the figure.
\begin{figure}[h]
\centerline{\psfig{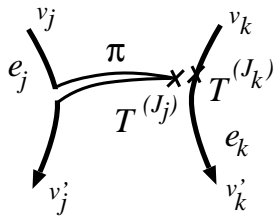}}
\caption{The basic configuration that defines the loop derivative.}
\label{loop}\end{figure}
Using (\ref{fierz}) one can rearrange the above expression in the
following way,
\begin{equation}
\Delta_{ab}(\pi_o^x) E(\raisebox{-5mm}{\psfig{file=der1.eps,height=11mm}},
\kappa) =  \kappa
\sum_{k} (-1)^{2(J_j+J_k)} \Lambda_{J_j J_k}  \epsilon_{abc}
\int_{e_k} dy^c \delta^3(x-y)
E\left(\raisebox{-5.7mm}{\psfig{file=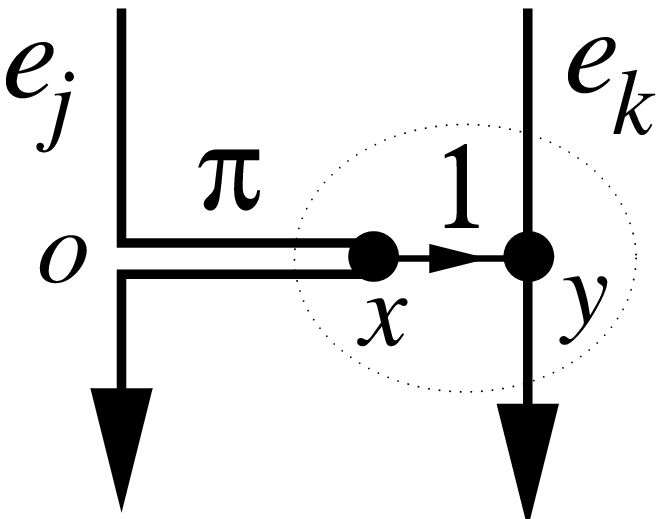,height=15mm}} \,\,,
\,\kappa\right).
\end{equation}
We now replace the two parallel lines in terms of a sum of
irreducible representations. The resulting expression can be cast
using recoupling identities\footnote{The recoupling relations
reduce the sum over the irreducible representations to a single
term with spin one. This is a difference with our
previous result -equation (28)- of reference \cite{GaGrPu98},
where we used a slightly different version of the Fierz identity.
The one we are using here is more useful in
calculations.} as,
\begin{equation}
\Delta_{ab}(\pi_o^x) E(\raisebox{-5mm}{\psfig{file=der1.eps,height=11mm}},
\kappa) =  \kappa
\sum_{k} (-1)^{2(J_j+J_k)} \Lambda_{J_j J_k}
\epsilon_{abc} \int_{e_k} dy^c \delta^3(x-y)
E\left(\raisebox{-5.9mm}{\psfig{file=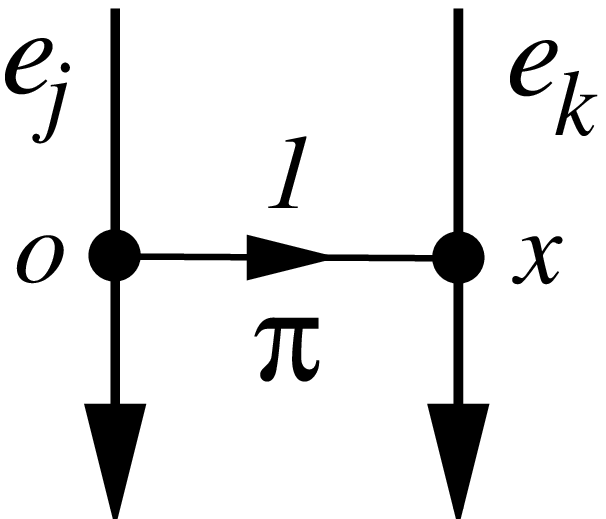,height=14.5mm}} \,\,,
\,\kappa\right).\label{deltae} 
\end{equation}

For future reference, we would like to introduce a formula for the
loop derivative when the path $\pi$ added is of infinitesimal
length. This is useful in the definition of the Hamiltonian
constraint. It could arise when the point $x$ is infinitesimally
separated along the same edge as the origin $o$ or it could correspond
to two infinitesimally adjacent edges (for instance incident on a
vertex). More precisely, in both the diffeomorphism and Hamiltonian
constraint one has the loop derivative evaluated along an
infinitesimal path multiplied times a regularized Dirac Delta
function. In that case one can rearrange the above expression in terms
of a chord diagram, 
\begin{equation}\label{deltavasi}
\lim_{\epsilon\rightarrow 0} \epsilon^2 
\int d^3x \chi_\epsilon(x,o)\Delta_{ab}(\pi_o^x) E(
\raisebox{-5mm}{\psfig{file=der1.eps,height=11mm}},\kappa)
=\lim_{\epsilon\rightarrow 0} \epsilon^2  
\kappa \sum_{k} \epsilon_{abc} \int_{e_k} dy^c
\chi_\epsilon(o,y)
E\left(\raisebox{-5.7mm}{\psfig{file=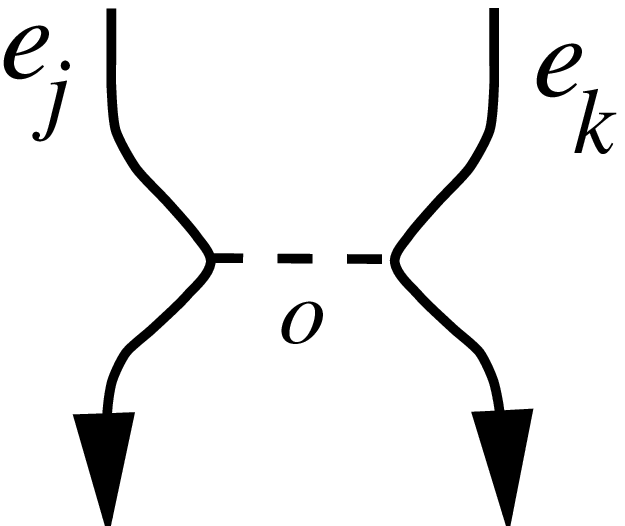,height=14.5mm}} \,\,,
\,\kappa\right), 
\end{equation}
where $\chi$ is a (symmetric, $\chi_\epsilon(x,y)=\chi_\epsilon(y,x)$) 
regulator such that $\lim_{\epsilon\rightarrow 0}
\chi_\epsilon(x,y) =  \delta^3(x-y)$. We will derive in the appendix the
rather remarkable result that the kind of derivative we considered in
the previous formula annihilates the framing-independent Vassiliev
invariants if the spin networks consider have trivalent or lower
intersections. This will allow us, in the companion paper, to find
large families of solutions to all the constraints of quantum gravity,
albeit with vanishing volume.

Notice that formula (\ref{deltae}) implies, if we analyze it order by
order in $\kappa$, that the loop derivative of a Vassiliev invariant
of order $n$ (the coefficient of $\kappa^n$ in the expansion) is
proportional to a Vassiliev invariant of order $n-1$.  That is, the
loop differentiation operation lowers the order of a Vassiliev
invariant by one.

As an instructive example of the consistency of the formalism, we can
compare the result predicted for the loop derivative of the first
order term in $\kappa$ by equation (\ref{deltae}) and the direct
computation of the loop derivative of expression (\ref{25}), 
\begin{equation}
\Delta^{(e)}_{ab}(\pi_o^x) <W(s)>^{(1)} =-{1 \over 2}
\Delta^{(e)}_{ab}(\pi_o^x) \left(\sum_{e_i,e_j \in s} r_{ij}
\int_{e_i}dy^c \int_{e_j} dz^d \epsilon_{cdf} \partial^f
{1 \over 4\pi |y-z|}\right)
\end{equation}
where we denoted that the loop derivative acts along the edge $e_i$ by
a superscript. The loop derivative will only act on the loop
dependence of the right-hand-side, i.e., on the integrals. To evaluate
this action, we recall the well known result (i.e. \cite{GaPubook}),
\begin{equation}
\Delta_{ab}^{(e)}(\pi_o^x) \int_e dy^c F_c(y) = \partial_{[a} F_{b]}(x),
\end{equation}
and therefore get,
\begin{equation}
\Delta^{(e_k)}_{ab}(\pi_o^x) <W(s)>^{(1)} = -
\sum_{e_j\in s} r_{kj}
\int_{e_j} dz^d \partial_{[a} \epsilon_{b]df} \partial^f
{1 \over 4\pi |x-z|}
\end{equation}
which we can rearrange as,
\begin{equation}
\Delta^{(e_k)}_{ab}(\pi_o^x) <W(s)>^{(1)} =
\sum_{e_j \in s} r_{kj}
\left(
\int_{e_j} dz^c \epsilon_{abc} \partial_f \partial^f
{1 \over 4\pi |x-z|}-
\int_{e_j} dz^f  \epsilon_{abc} \partial_f \partial^c
{1 \over 4\pi |x-z|}\right).
\end{equation}
The last term gives the integral along an edge of a quantity
differentiated with respect to the integration variable. This gives
contributions only at the ends of the edges, and it gives the same
contribution per each incoming edge. Recalling that the summation of
all $r_{ij}$'s for each vertex vanishes, the last term vanishes
when added along the whole spin network, vertex per vertex. The
remaining term gives the final result,
\begin{equation}
\Delta^{(e_k)}_{ab}(\pi_o^x) <W(s)>^{(1)} =
\sum_{e_j\in s} r_{kj}
\epsilon_{abc} \int_{e_j} dy^c \delta^3(x-y).
\label{deltagauss}
\end{equation}
This expression coincides with the right hand side of (\ref{deltae})
evaluated to first order in $\kappa$. The group factors $r_{ij}$
are given in (\ref{deltae}) by the chromatic evaluation of the
invariant $E$, which is what one gets at zeroth order in $\kappa$.

It is worthwhile emphasizing that although expressions like
(\ref{deltae}) can be viewed as ``formal manipulations'' in the path
integral that one uses to motivate certain results, yet the
perturbative Feynman-diagrammatic calculations yield order by order
results that are consistent with the formal manipulations. We will not
discuss in detail similar calculations for invariants of higher order,
but they can be checked in detail using the formulae for loop
derivatives of multitangents, see for instance \cite{GaPubook}.

An interesting observation is that if we go back to expression
(\ref{tata}) we see that the loop derivative introduces in the spin
network two Lie algebra generators contracted. This is precisely the
same contribution we got in (\ref{vasivasi}) for the Vassiliev
derivative. This points out to a very appealing connection between the
loop derivative and the Vassiliev derivative. It therefore highlights
why Vassiliev invariants play a special role among knot invariants
from the point of view of quantum gravity: they are naturally
``differentiable'' knot invariants, and the derivative in question is
nothing else but the loop derivative. This is consistent with the fact
that the loop derivative ``lowers the order'' of a Vassiliev
invariant, as we saw before in the example of the invariant of order
one being related to the zeroth order invariant. Freidel was the first
to notice, and is currently studying the implications of this
connection \cite{Fr}.

\subsection{The diffeomorphism constraint and the loop derivative}

To define a diffeomorphism constraint we will formally integrate by
parts the expression of the constraint in the connection
representation acting on a state given by the loop transform. In terms
of the new variables, the diffeomorphism constraint is given
classically by $C(\vec{N}) = \int d^3x N^a(x) \tilde{E}^b_i
F_{ab}^i$. If one wishes to promote this to a quantum operator one
needs to choose a factor ordering and a regularization. We will choose
the factor ordering in which the triad is at the left. In this factor
ordering, the operator formally fails to generate diffeomorhpisms on
the wavefunctions in the connection representation. This can be
amended if one regularizes the operator by point-splitting the
operator $E$ and $F$ and using a symmetric regularization
\cite{BrGaPunpb}. We will therefore operate with such an operator on
a state in the loop representation defined via the loop transform,
\begin{eqnarray}
C(\vec{N}) \Psi(s) &=& \int DA (C(\vec{N}) \Psi(A)) W_A(s) \\&&= 
\int DA \int d^3x \int d^3y
\lim_{\epsilon\rightarrow 0} {(N^a(x)+N^a(y))\over 2}
\chi_\epsilon(x,y) \left({\delta \over \delta A_b^{\alpha}(x)} F_{ab}^{\alpha}(y)
\Psi(A)\right) 
W_A(s),\nonumber
\end{eqnarray}
where again $\chi$ is a (symmetric, $\chi_\epsilon(x,y)=\chi_\epsilon(y,x)$) 
regulator such that $\lim_{\epsilon\rightarrow 0}
\chi_\epsilon(x,y) =  \delta^3(x-y)$,  and the symmetric
regularization refers to taking the symmetric average of the vector
$\vec{N}$ at the points $x$ and $y$. This expression is not gauge
invariant due to the point splitting. This can be fixed by introducing
small pieces of holonomies along paths $\pi_y^x$ that connect the
points $x$ and $y$ and inserting them to get a gauge invariant
expression. 
If one now formally integrates by parts the functional derivative and 
represents the $F_{ab}$ using the loop derivative, we get,
\begin{equation}
C(\vec{N}) \Psi(s) = \sum_{k}
\int DA \Psi(A) 
\lim_{\epsilon\rightarrow 0} \int d^3x \int_{e_k} dy^b
{(N^a(x)+N^a(y))\over 2}\chi_\epsilon(x,y) \Delta_{ab}(\pi_y^x) W_A(s).
\label{diffwil}
\end{equation}
We therefore identify the following operator as the infinitesimal
generator of diffeomorphisms in terms of spin nets,
\begin{equation}
C(\vec{N}) \Psi(s) = \sum_{k}
\lim_{\epsilon\rightarrow 0} \int d^3x \int_{e_k} dy^b
{(N^a(x)+N^a(y))\over 2}\chi_\epsilon(x,y) \Delta_{ab}(\pi_y^x)
\Psi(s).
\end{equation}

The last expression should be understood as a shorthand notation for 
(\ref{diffwil}). This means that the limit should be taken before
evaluating the functional integral, otherwise we would fall in the
usual pitfall of not having a defined action for a loop derivative on
a knot invariant. This action {\em is defined} by formula
(\ref{diffwil}) for the invariants in question, replacing $\Psi(A)$ by
the exponential of the Chern--Simons form. The last expression has
been well known for some time to correspond to the infinitesimal
generator of diffeomorphisms on Wilson loops evaluated along smooth
loops. In the case of spin networks, the expression displaces each
edge infinitesimally. The vertices are left fixed, but are reconnected
to the shifted edges by retraced paths. Due to the gauge invariance of
the intertwiner operators, the functions behave as if the vertex had
been shifted along with the edges.

In our previous work in terms of loops \cite{GaPubook,GaGaPu} we have
used an (unregulated) diffeomorphism constraint that looked slightly
different than the one we consider here, but that coincides in the
limit, when acting on functions on which the loop derivative yields a
smooth result, like for instance holonomies of smooth connections. The
main difference is that in the regulated operator the path $\pi$
extends out of the original loop and tends to it in the limit, whereas
in the unregulated expressions one simply took the path directly along
the loop.

The formula (\ref{diffwil}) involves evaluating the loop derivative
for an infinitesimal path. To evaluate this, it is convenient to go
back to (\ref{tata}), and notice that the addition of the two Lie
algebra generators in the two nearby points corresponds to a chord
diagram in which a dashed line starts and ends in a given edge of the
spin network. Another possibility is if the action is close to a
vertex. In that case one can have the insertion of the two Lie algebra
generators in two different links adjacent to the vertex. Concretely,
we get,
\begin{eqnarray}
&&C(\vec{N}) E\left(
\raisebox{-5mm}{\psfig{file=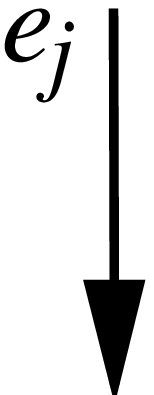,height=11mm}}\, , \kappa\right)
= -{\kappa \over 2} \lim_{\epsilon\rightarrow 0}  \int_{e_j} dy^b
\int_{e_j} dx^c ({N^a(y) + N^a(x)\over 2})\epsilon_{abc} \chi_\epsilon(x,y)
E\left(\raisebox{-5.5mm}{\psfig{file=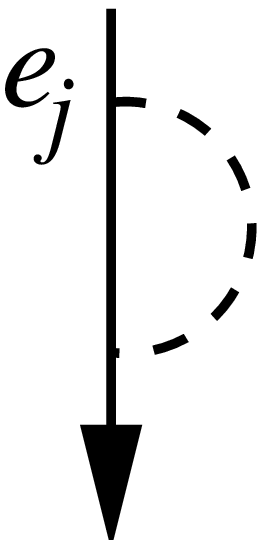,height=13mm}}\, ,
\kappa\right) =0\label{writh1}\\
&&C(\vec{N}) E\left(
\raisebox{-5mm}{\psfig{file=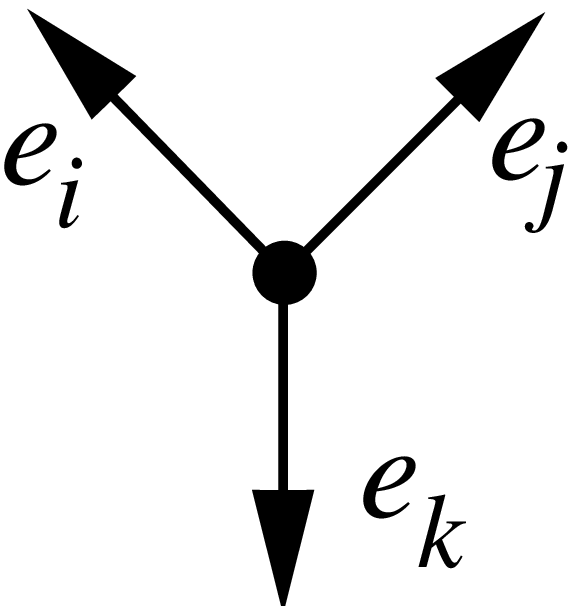,height=11mm}}\, , \kappa\right) =
0
\label{writh2}
\end{eqnarray}
where we see that the contributions vanish in the first expression due
to the contraction of the $\epsilon_{abc}$ with a symmetric expression
in $b,c$. In the second expression, given the action on an edge $i$,
the contribution arising from connecting that edge with another edge
$j$ is equal and opposite to that arising from considering the action
on the edge $j$ and considering the connection with the edge $i$, and
they therefore cancel each other as well. This is all only true
because we have chosen a symmetric regularization, otherwise the
expression would not be symmetric under the interchange of $b$ and
$c$.

We should notice that this is a major departure from what we did in
our previous paper \cite{GaGrPu98}. There we did not choose the
symmetric regularization, and the diffeomorphisms did not vanish on
these invariants. At that point we interpreted the action of the
constraint as associated with the change in the value of the framing
invariants due to the addition of ``writhes'' by the diffeomorphism.
It should be noted that there is some ambiguity in what one means by
diffeomorphism in the case of framing dependent invariants, i.e., if
one is talking of diffeomorphisms of ``ribbons'' or of loops with no
width. Choosing the invariants to be diffeomorphism invariant is in
line with adopting the point of view that one is considering
diffeomorhpisms of ``ribbons''.

What we find here is that the motivation for the symmetric
regularization that came from the connection representation has a
natural counterpart in loops: in the connection representation the
regularization ensured that the constraint generate diffeomorphisms,
and therefore that the Chern--Simons state be annihilated by the
constraint. The same is true in the loop representation. With this
regularization, the loop transforms of that state are diffeomorphism
invariant. In other words, in this regularization, the diffeomorphism
considers the framing dependent objects as invariants. All
coefficients in the expansion of the expectation value of the Wilson
loop, and in particular all the independent components of each
coefficient (the Vassiliev invariants) we discussed before, are
annihilated by the diffeomorphism constraint.  That is, in the
regularization chosen, both the framing independent and the framing
dependent invariants are annihilated by the diffeomorphism
constraint. It is still open to question if the use of
framing-dependent objects is warranted in quantum gravity, where the
whole formalism has been from the outset set up in terms of loops
without any reference to a framing.

\section{Conclusions}

To conclude, we summarize on the results found up to now: a) We have a
space of wavefunctions of spin networks that are loop differentiable
(the Vassiliev invariants and all the coefficients of the expansion of
the expectation value of the Wilson loops); b) The diffeomorphism
constraint naturally constructed in terms of the loop derivative is
well defined and annihilates these states.

We have elucidated several aspects of the generalization of Vassiliev
invariants to spin networks. To begin with, we see that in general the
expectation value of a Wilson net cannot be factorized as a framing
independent invariant times a framing dependent prefactor as was the
case for loops. This, however, is possible if the spin networks have
trivalent or simpler intersections. It nevertheless possible to
identify at each order in the perturbation expansion, and for a given
spin network, invariant quantities that are ambient isotopic (framing
independent) and we illustrate the procedure to find them up to
second order in the perturbative expansion. We have given expressions
for the invariants that are given by analytic integrals along the
lines that form the spin network. This allows to operate with these
wavefunctions in a direct manner with the operators we will discuss in
the following paper.

The fact that the operator we chose for the diffeomorphism constraint
generates the geometrical action of diffeomorphisms on Wilson nets,
suggests that the algebra of constraints will reproduce the expected 
classical Poisson structure. At the moment however, we have only
constructed states in the loop representation where the constraint
vanishes and therefore the algebra is reproduced, albeit trivially. In 
the companion paper we will introduce ``habitats'' of
non-diffeomorphism invariant states related to the ones we introduced
here and we will explicitly show that the classical Poisson algebra
is reproduced at the quantum level.

Having a well defined action for the loop derivative opens the
possibility of introducing quantum versions of the Hamiltonian
constraint and to implement in a concrete well defined setting
proposals for the Hamiltonian constraint that have been shown to have
the correct algebra at a formal quantum level \cite{GaGaPu}. We will
expand on this and other topics in the companion paper.

\acknowledgements We wish to thank Abhay Ashtekar, Laurent Freidel,
 and Thomas Thiemann for comments and discussions.
 This work was supported in part by the National Science Foundation
 under grants NSF-INT-9811610, NSF-PHY-9423950, NSF-PHY-9407194,
 research funds of the Pennsylvania State University, the Eberly
 Family research fund at PSU.  JP acknowledges support of the Alfred
 P. Sloan and John Simon Guggenheim foundations. We acknowledge
 support of PEDECIBA (Uruguay). RG and JP wish to thank the
 Institute for Theoretical Physics of the University of California at
 Santa Barbara and CDB, RG and JG the Center for Gravitational Physics
 and Geometry at Penn State for hospitality during the completion of
 this work.

\appendix
\section{Loop derivative of framing independent Vassiliev invariants}

In this appendix we will prove a property of a limiting behavior of
the loop derivative of the framing independent Vassiliev invariants
based on {\em trivalent} intersections. What we will show is that in the
limit in which the path on which the loop derivative shrinks to zero,
the loop derivative vanishes. The interest of this limit is that it is
the one that appears in the loop derivatives that arise in the
diffeomorphism and Hamiltonian constraints of quantum gravity in the
loop representation that we will discuss in the forthcoming paper.

To prove this statement, we start by defining the invariant $J(s,\kappa)$,
\begin{equation}
E(s,\kappa)\equiv \exp\left[{\kappa \over E(s,0)} <W(s)>^{(1)}\right]
J(s,\kappa), 
\label{jones}
\end{equation}
that is, we extract the framing dependence in the prefactor, something
that is allowed for trivalent or lower intersections. 

Using the recoupling identities for trivalent intersections and
equations  (\ref{deltavasi}) and (\ref{deltagauss}), one gets,
\begin{equation}
\lim_{\epsilon\rightarrow 0} \epsilon^2
\int d^3y \chi_\epsilon(y,o) \Delta_{ab}(\pi_o^y) E(s,\kappa) = 
\lim_{\epsilon\rightarrow 0} \epsilon^2
E(s,\kappa) \int d^3y \chi_\epsilon(y,o) \Delta_{ab}(\pi_o^y) 
\left[ {\kappa \over E(s,0)} <W(s)>^{(1)}\right],
\end{equation}
and comparing this expression with the derivative of (\ref{jones}),
and bearing in mind that the loop derivative satisfies Leibnitz' rule,
one immediately concludes that
\begin{equation}
\lim_{\epsilon\rightarrow 0} \epsilon^2 
\int d^3y \chi_\epsilon(y,o) \Delta_{ab}(\pi_o^y) J(s,\kappa)=0
.
\end{equation}
Since the result is independent of $\kappa$, it implies that all terms
in the expansion of $J(s,\kappa)$ in powers of $\kappa$ is annihilated
by the loop derivative as well. Moreover, since we could have carried
this construction for {\em any gauge group} (the only elements we used
was that recoupling identities for trivalent intersections are
tantamount to extracting a prefactor, which is true for any group), we
conclude that the topological invariants are annihilated by this limit
of the loop derivative.

\end{document}